\documentclass[onecolumn,authoryear]{els-mrw} 

\usepackage{amsmath,amssymb,amsfonts,amsthm,makeidx,graphicx}
\usepackage{txfonts}
\usepackage{helvet}
\usepackage{wrapfig}
\usepackage{cite}
\usepackage{url}

\begin{document}

\chapter{Tidal interactions in stellar and planetary systems}\label{chap1}

\author[1]{Adrian J. Barker}
\address[1]{\orgname{School of Mathematics, University of Leeds,} \orgaddress{Leeds, LS2 9JT, UK}}

\articletag{Chapter Article tagline: update of previous edition,, reprint..}

\maketitle

\begin{abstract}[Abstract]
Gravitational tidal interactions drive long-term rotational and orbital evolution in planetary systems, in multiple (particularly close binary) star systems and in planetary moon systems. Dissipation of tidal flows in Earth's oceans is primarily responsible for producing gradual expansion of the Moon's orbit at a few centimetres per year as the Earth's day lengthens by a few milliseconds per century. Similar processes occur in many astrophysical systems. For example, tidal dissipation inside (slowly rotating) stars hosting short-period planets can cause the orbits of these planets to decay, potentially leading to planetary destruction; tidal dissipation inside stars in close stellar binary systems -- and inside short-period planets such as hot Jupiters in planetary systems -- can cause initially eccentric orbits to become circular. To model these processes, explain many current observational results, and make predictions for future observations, we require a detailed theoretical understanding of tidal flows and the mechanisms by which -- and how efficiently -- they are dissipated inside stars and planets. This article will introduce our current understanding of tidal flows and dissipation inside stars (and to a lesser extent giant planets), as well as highlight some unsolved problems.
\end{abstract}

\begin{keywords}
tides, tidal dissipation, tidal evolution, tidal interactions, tidal flows, tidal deformations
\end{keywords}

\begin{glossary}[Glossary]
\term{Tidal gravity}: The differential gravitational potential felt across an extended body such as a star or planet due to another body. \\
\term{Tidal deformation}: The large-scale deformation of a star or planet (``tidal bulges") by the tidal potential of another body. \\
\term{Tidal flow}: The response of a gaseous (or liquid) body to the tidal potential of another body, consisting of both large-scale flows that carry the bulge around and waves. \\
\term{Tidal frequency}: A frequency at which fluid in a star or planet is forced by the tidal potential. \\
\term{Tidal dissipation}: The dissipation of time-dependent tidal flows inside a star or planet, which drives spin-orbit evolution. This is likely to be dependent on tidal frequency, amplitude and stellar properties (mass and age, i.e., structure, and rotation). \\
\term{Tidal quality factor}: An inverse measure ($Q$, or the modified version $Q'$) of the efficiency of tidal dissipation defined by analogy with the quality factor of a forced, damped, harmonic oscillator. Larger values imply relatively inefficient dissipation, smaller values imply relatively more efficient dissipation. It is not generally a constant parameter for a given star, and can vary substantially depending on tidal mechanism, frequency and stellar properties.\\
\term{Spin-orbit evolution}: Evolution of the axial rotations (spins) of a star or planet, as well as evolution of the orbital elements, including semi-major axis, eccentricity (a measure to quantify how elliptical the orbit is), and various angles indicating orientation of the orbit relative to both a reference direction and to axial rotations. \\
\term{Spin-orbit synchronisation}: The tidal evolution of a star's (or planet's) axial angular rotation rate to match the angular frequency of its (or its companion's) orbit, which is sometimes referred to as tidal locking. \\
\term{Orbital circularisation}: The decay of a body's orbital eccentricity ($e<1$) from a non-zero value towards $e=0$ (circular orbit), which can be caused by tidal dissipation in stars that start out on eccentric (elliptical) orbits.\\
\term{Circularisation period}: Essentially, this is the largest orbital period for which binary stars (or hot Jupiters) are primarily circular, with those orbiting with longer periods tending to be eccentric.\\
\term{Hot Jupiter}: a giant gaseous planet similar in mass similar to (or somewhat larger than) Jupiter in our own Solar System orbiting its star in less than approximately 10 Earth days.
\end{glossary}

\begin{BoxTypeA}[keypts]{Key Points}
\begin{itemize}
\item Tidal evolution is important in close stellar systems, and in planetary systems, particularly those hosting close-in planets -- and it can significantly modify the rotations (spins) and orbits of planets and stars, typically over millions to billions of years.
\item Tidal evolution is driven by the dissipation of tidal flows inside planets and stars. If we know how efficiently tidal flows are dissipated, we can infer the orbital and rotational evolution that this will cause. This motivates theoretical studies to determine the mechanisms of tidal dissipation in stars and planets to explain current observations, as well as observational studies to constrain tidal theories.
\item A variety of fluid dynamical mechanisms are thought to be responsible for tidal dissipation in both convection and radiation zones; which one dominates is predicted (and inferred from observations) to vary depending on the particular scenario, and on the stellar and planetary properties.
\item This is an evolving field, with much observational and theoretical progress expected over the next few years, partly with space missions such as the upcoming PLATO mission, as well as ongoing ground-based studies.
\end{itemize}
\end{BoxTypeA}

\section{Introduction: what are tides and why are they important?}\label{chap1:sec1}

Gravitational tidal interactions play crucial roles in many areas of astrophysics, including the evolution of close binary stars and planetary systems hosting short-period planets. They will also determine the ultimate fate of the Earth when the Sun becomes a red giant. Some aspects of tidal interactions are well understood and some are not. This article will introduce the problem and briefly summarise the state of our knowledge of tidal flows in stars and, to a lesser extent, giant planets, which are bodies that are predominantly fluid.

\begin{figure}[b]
\centering
\includegraphics[width=0.6\textwidth,trim=4cm 15cm 10.5cm 0cm,clip=true]{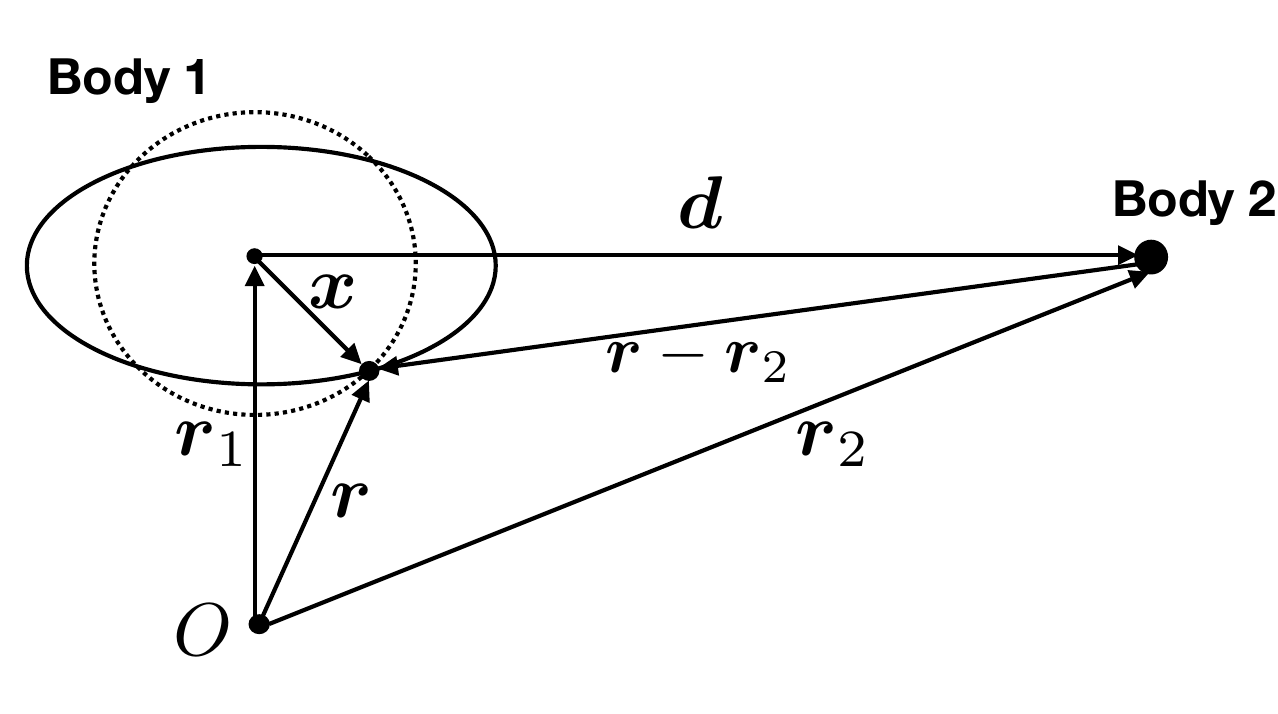}
\caption{Diagram to illustrate gravitational tidal interactions between a body 1 (the primary star or planet) that is subjected to the tidal gravity of body 2 (modelled here as a point mass, without structure).}
\label{chap1:fig1}
\end{figure}

To set the scene, consider a fluid body of mass $M_1$ of unperturbed radius $R_1$ rotating at the rate $\Omega_s$ about its spin axis, such as a star or gaseous planet, which is perturbed by a point-mass companion (that could represent a star, planet, or even a black hole, but we will ignore its structure) of mass $M_2$. We define an arbitrary origin $O$ and specify the position vectors of the centres of each body with respect to this as $\boldsymbol{r}_1$ and $\boldsymbol{r}_2$, respectively, as shown in Fig.~\ref{chap1:fig1}. Both bodies orbit the centre of mass of the system and the orbital separation between both bodies (as a function of time $t$) is $\boldsymbol{d}(t)=\boldsymbol{r}_2(t)-\boldsymbol{r}_1(t)$. We define $\boldsymbol{x}=\boldsymbol{r}-\boldsymbol{r}_1$ as the position vector between the centre of body 1 and a chosen point in its interior or surface with position vector $\boldsymbol{r}$. The gravitational potential\footnote{In Newtonian gravity. Corrections due to General Relativity are usually very small when computing tidal responses of star and planets \citep[though see, e.g.,][]{Andersson2021}, though they can produce precession of eccentric orbits, for example.} due to body 2 can be Taylor expanded about the centre of body 1 assuming $|\boldsymbol{x}|\ll |\boldsymbol{d}|$, relevant when the orbital separation is much larger than the radius of body 1, to obtain
\begin{align}
\label{Phi}
\Phi (\boldsymbol{x},t) = -\frac{GM_2}{|\boldsymbol{r}-\boldsymbol{r}_2|} = -\frac{GM_2}{|\boldsymbol{x}-\boldsymbol{d}|} =  -\frac{GM_2}{|\boldsymbol{d}|}\left(\underbrace{1}_{\text{no contribution to forces}}+\underbrace{\frac{\boldsymbol{d}\cdot\boldsymbol{x}}{|\boldsymbol{d}|^2}}_{\text{produces orbital motion}}+\underbrace{\frac{3(\boldsymbol{d}\cdot\boldsymbol{x})^2-|\boldsymbol{x}|^2 |\boldsymbol{d}|^2}{2|\boldsymbol{d}|^4}+O\left(\frac{|\boldsymbol{x}|^3}{|\boldsymbol{d}|^3}\right)}_{\equiv \Psi (\boldsymbol{x},t)\; \text{(tidal potential)}}\right).
\end{align}
Remember that the gravitational force on a mass $m$ due to the potential $\Phi$ is computed by taking (minus) the gradient (with respect to $\boldsymbol{x}$) of the potential and multiplying by the mass, i.e., the force is $-m\nabla \Phi$. The first term on the right hand side is independent of $\boldsymbol{x}$ and provides no contribution to the force as it has vanishing gradient. The second term produces a spatially-constant force that produces orbital motion of body 1 about the centre of mass and would be present even if body 1 was a point-mass. The third term, and any subsequent terms, constitute the tidal potential $\Psi$, and denote the variation in the gravitational potential of body 2 over body 1. This is responsible for producing tidal deformations and exciting tidal flows within body 1, which when dissipated contribute to orbital and rotational evolution. 

It is appropriate for slow tidal evolution, and also convenient, to assume the orbit to be Keplerian at any given instant and to introduce its orbital elements: $a$ is semi-major axis ($|\boldsymbol{d}|=a$ for circular orbits), $e$ is eccentricity, $i$ is inclination or obliquity of body 1 -- the angle between its rotation axis and the orbit normal vector -- and $\Omega_o=\sqrt{G(M_1+M_2)/a^3}=2\pi/P$ is the orbital mean motion (where $G$ is the gravitational constant) -- exactly equal to the angular frequency of the orbit if it is circular, and $P$ is the orbital period. 

\subsection{Tidal potential and deformation}
\label{TidalPot}

Let us first consider the simplest case of a circular, aligned orbit ($e=i=0$ and $|\boldsymbol{d}|=a$) and a non-synchronous rotation $\Omega_s\ne \Omega_o$. This is, for example, relevant for tides in the Earth due to the Moon because the Earth rotates much faster (with a rotation period of one day) than the Moon orbits the centre of mass of the Earth-Moon system (with an orbital period of a month). We may define the orbital plane (which is also the equatorial plane of body 1) to coincide with the $(x,y)$-plane, so that the rotation axis (and orbit normal vector) is along $z$, using Cartesian coordinates. We introduce spherical polar coordinates $(r,\theta,\phi)$ centred on body 1 and in which $\theta=0$ lies along $z$, so that $\boldsymbol{x}=(x,y,z)=r(\sin\theta\cos\phi,\sin\theta\sin\phi,\cos\theta)$ and the Cartesian coordinates of body 2 at a given time depend on its orbital angle $M(t)$, so that $\boldsymbol{d}(t)=a (\cos M (t),\sin M(t),0)$ since body 2 is orbiting body 1 and the reference frame is not rotating. The largest component (with $l=2$, ignoring any larger $l$ components for now; $l$ will be defined below) of the tidal potential in Eq.~\ref{Phi} is then
\begin{align}
\Psi(\boldsymbol{x},t)=\frac{GM_2 r^2}{2 a^3}\left(1-3\sin^2\theta\cos^2 (\phi-M(t))\right)=\frac{GM_2 r^2}{4 a^3}\left(2-3\sin^2\theta \, \left(1+\cos (2\phi-2\Omega_o t)\right) \right),
\end{align}
since $M(t)=\Omega_o t$ for a circular orbit if $M(t=0)=0$, and we have employed the double angle formula $2 \cos^2 \varphi = 1+\cos 2\varphi$. This contains both static and time-dependent parts, where the former just causes a deformation of body 1 and the latter causes both a time-dependent deformation and an associated flow inside the body, which is the component we will primarily focus upon in this article. It is conventional to express the tidal potential using spherical harmonics ($Y_l^m(\theta,\phi)$, where $l$ is the integer degree and $m$ the integer azimuthal order; e.g.~Section 18.3 of \citealt{RHB2006}), in which case we may write\footnote{The real part must be taken because we live in a real world and not an imaginary one, as far as we know.}
\begin{align}
\label{Psi1}
\Psi(\boldsymbol{x},t) = \frac{GM_2 r^2}{a^3}\mathrm{Re}\left\{\underbrace{\sqrt{\frac{\pi}{5}}Y_2^0(\theta,\phi)}_{\text{static tide}} -\underbrace{\sqrt{\frac{6\pi}{5}} Y_2^2(\theta,\phi)\mathrm{e}^{-2\mathrm{i} \Omega_o t}}_{\text{asynchronous tide}}\right\},
\end{align}
where we have introduced the spherical harmonics $Y_2^0(\theta,\phi)=\frac{1}{4}\sqrt{\frac{5}{\pi}}(3\cos^2\theta-1)$, which is independent of $\phi$, $Y^2_2(\theta,\phi)=\frac{1}{4}\sqrt{\frac{15}{2\pi}}\sin^2\theta \,\mathrm{e}^{2\mathrm{i}\phi}$, and we remember both Euler's/de Moivre's formula $ \mathrm{e}^{\mathrm{i}m\phi}=\cos m\phi+\mathrm{i}\sin m\phi$, and $\cos^2\theta+\sin^2\theta=1$. In the reference frame rotating with body 1 at the rate $\Omega_s$ about $z$, the tidal potential can be expressed exactly as in Eq.~\ref{Psi1} except that we must replace the frequency $2\Omega_o$ with the Doppler-shifted one (and coordinates are now defined in the rotating frame), which we shall define as $\omega_{2,2,2}=2\Omega_o-2\Omega_s$. This is the frequency at which rotating fluid in body 1 is forced by the time-dependent parts of the potential $\Psi$ (the asynchronous tide here). This case is, for example, relevant for Earth tides due to the Moon, and the factor of 2 here simply represents having two high tides and two low tides in each day (approximately, because $\Omega_s\gg \Omega_o\ne 0$). The key contributions to the amplitude of the tidal potential in Eq.~\ref{Psi1} are given by the factor
\begin{align}
\label{PsiAmp}
\frac{GM_2r^2}{a^3} = \epsilon_T \omega_d^2 r^2,
\end{align}
where $\omega_d=\sqrt{GM_1/R_1^3}=2\pi/P_d$ is called the dynamical frequency of body 1 (with associated dynamical period $P_d\approx 2.7 $ hours for the Sun) and is similar to the lowest frequency surface gravity oscillations (usually referred to as f-modes), and the dimensionless tidal amplitude in body 1 is 
\begin{align}
\epsilon_T = \left(\frac{M_2}{M_1}\right)\left(\frac{R_1}{a}\right)^3=\frac{M_{2}}{M_1+M_2}\left(\frac{P_d}{P}\right)^2.
\end{align}
The response of body 1 to the potential $\Psi$ is discussed in more detail in \S~\ref{EqmTide} and \ref{DynTide}, but we will briefly illustrate it here.  If body 1 instantaneously adjusts to the time-varying potential in a quasi-hydrostatic manner (which we refer to as the equilibrium tide), it will be deformed by each component of the potential $\Psi$, with a radial displacement of fluid elements from their unperturbed configuration of the form
\begin{align}
\xi_r(r,\theta,\phi,t)\propto \epsilon_T R_1 \mathrm{Re}\left\{ f(r) Y_2^2(\theta,\phi)\,\mathrm{e}^{-\mathrm{i}\omega_{2,2,2}t} \right\},
\label{xir}
\end{align}
due to the $l=m=2$ component of Eq.~\ref{Psi1} in the rotating frame, where $f(r)$ is a (dimensionless) function of radius that can be computed (see \S~\ref{EqmTide}). There is a similar, but static, deformation corresponding to the $l=2,m=0$ component in Eq.~\ref{Psi1}. The deformation described by Eq.~\ref{xir} has angular structure $\propto \cos (2\phi-\omega_{2,2,2}t)$, and the surface of the body is deformed according to $R_1+\xi_r(R_1,\theta,\phi,t)$; therefore, going round in azimuth $\phi$ at each time $t$, there are two high tides (where $\xi_r$ is positive, corresponding to the ``tidal bulges") and two low tides (where $\xi_r$ is negative), and this (prolate spheroidal, pointing towards the companion) deformation propagates azimuthally at the frequency $\mathrm{d}\phi/\mathrm{d} t=\omega_{2,2,2}/2=\Omega_o-\Omega_s$. The latitudinal structure is $\propto \sin^2 \theta$, describing the tidal bulges lying in the plane containing body 2. The corresponding tidal flow (i.e.~velocity field, given by the time derivative of Eq.~\ref{xir}) required to move around these tidal bulges is illustrated later in Fig.~\ref{chap1:fig3}.

When $\epsilon_T\ll 1$, which is typically the case\footnote{For example, consider a very short-period binary system containing two stars exactly like the Sun orbiting their centre of mass in $P=1$ day. This would have $\epsilon_T=6.67\times 10^{-3}$, whereas a wider binary with $P=10$ days would have $\epsilon_T=6.67\times 10^{-5}$ instead.}, it means the overall tidal deformation of the star, with surface radial displacement (from the unperturbed body) $\xi_r\propto \epsilon_T R_1$, is small, so we can approximate the problem by studying the response of a spherical body to a small tidal perturbation (oblate spheroidal about the rotation axis if it is rotating rapidly such that $\Omega_s^2/\omega^2_d\nll 1$). While small values of $\epsilon_T$ suggest that the tide may be modelled well using linear tidal theory (which we implicitly employed to write down Eq.~\ref{xir}), this is only part of the story, because tidal waves can have much shorter wavelengths than the large-scale deformation, and these waves can be subject to important nonlinear effects for much smaller amplitudes, even when $\epsilon_T\ll 1$. We will discuss this further in \S~\ref{mechanisms}.

When $\epsilon_T\gtrsim 1$ (for which the tide is in a highly nonlinear regime), the tidal gravitational force near the surface of body 1 due to body 2 ($\approx GM_2 R_1/a^3$) exceeds body 1's self-gravity ($\approx GM_1/R_1^2$). Hence, we expect stars or planets to be tidally disrupted and break up if $\epsilon_T\gtrsim 1$, so that their orbital separations satisfy
\begin{align}
a\lesssim a_R \approx \left(\frac{M_2}{M_1}\right)^{\frac{1}{3}}R_1.
\end{align}
This is referred to as the ``Roche limit" and it depends on the mass ratio \citep[e.g.][]{Roche}. If we have two identical bodies, we find $a_R\approx R_1$, whereas if body 2 is a supermassive black hole with $M_2\approx 10^6M_\odot\approx 2\times 10^{36} \mathrm{kg}$, $a_R\approx 100 R_1$. Tidal disruption events are transient astronomical sources in which a star passes so close to a supermassive black hole that the star is destroyed by the tidal gravitational forces of the latter \citep[e.g.][]{Hills1975,Rees1988,TDE2021}. In reality, $a_R$ is slightly larger than this simple estimate (so that bodies are destroyed for slightly wider separations) by a constant factor that depends on the nature of the body. In perfectly rigid planetary bodies, this factor is approximately 1.26 and in the extreme limit of a homogeneous (constant density) fluid body it is approximately 2.45, with realistic bodies lying somewhere in between \citep[e.g.][]{C1969,Faber2005,Guill2013}. We will primarily focus upon $\epsilon_T\ll 1$ in the rest of this article.

In the general case with an arbitrarily eccentric ($e\ne 0$) and misaligned ($i\ne 0$) orbit, the tidal potential is typically expressed in a similar way to Eq.~\ref{Psi1}, as a series of spherical harmonics ($Y_l^m(\theta,\phi)$) and temporal Fourier modes (frequencies) as\footnote{Note that $\nabla^2\Psi=0$ because it is a component of the gravitational potential of body 2 outside that body, where its density vanishes (so Poisson's equation reduces to Laplace's equation). Hence, each component must have a spatial form $\propto r^l Y^m_l(\theta,\phi)$; see section 4.3 of \citet{MD1999} or Section 21.3.1 of \citet{RHB2006}. The governing equation is linear so we can sum up linearly independent components. The trickiest part of this expression is determining the coefficients $A_{l,m,n}(e,i)$.} \citep[e.g.][]{Kaula1961,PS1990,Ogilvie2014}
\begin{align}
\Psi(\boldsymbol{x},t)=\Psi  (r,\theta,\phi,t)= \mathrm{Re}\left\{ \sum_{l=2}^{\infty}\sum_{m=0}^{l}\sum_{n=-\infty}^{\infty}\frac{GM_2}{a} A_{l,m,n} (e,i) \left(\frac{r}{a}\right)^l Y_l^m(\theta,\phi)\mathrm{e}^{-\mathrm{i}\omega_{l,m,n}t}\right\},
\label{Psi}
\end{align}
again using spherical polar coordinates $(r,\theta,\phi)$ centred on body 1 and with its rotation axis along $\theta=0$. We have chosen to work in the reference frame that rotates with body 1 at the rate $\Omega_s$, which means that the tidal frequency of each component in the non-rotating frame $n\Omega_o$ (where $n$ is an integer) is Doppler-shifted depending on its azimuthal wavenumber $m$ according to $\omega_{l,m,n}=n\Omega_o - m \Omega_s$. The tidal frequency $\omega_{l,m,n}$ represents the frequency at which fluid is forced in body 1 by the $(l,m,n)$ component of the tidal potential. For the important case derived above in Eq.~\ref{Psi1} for the case of a non-synchronised ($\Omega_s\ne \Omega_o$) rotating body on a circular, aligned (equatorial, with $i=0$) orbit the relevant tidal component is $l=m=n=2$ with frequency $\omega_{2,2,2}=2(\Omega_o-\Omega_s)$. In the most general configuration there can be many relevant tidal components with different frequencies. This is perhaps easiest to show in the case of a weakly eccentric orbit with $e\ll 1$, where it can be shown that $|\boldsymbol{d}|\approx a(1-e\cos \Omega_o t)$ and $M\approx \Omega_o t + 2 e \sin \Omega_o t)$, so instead of Eq.~\ref{Psi1}, $\Psi$ would contain components with integer multiples of $\Omega_o$, leading to four time-dependent components (one with $m=0$ and three with $m=2$) in the rotating frame \citep[e.g.][]{Zahn1977,OL2004}.

In Eq.~\ref{Psi} there are a set of coefficients $A_{l,m,n}(e,i)$ for each $(l,m,n)$ component that depend on the eccentricity $e$ and inclination/obliquity $i$. For a circular ($e=0$) and aligned orbit ($i=0$), as we derived in Eq.~\ref{Psi1}, there are only two non-vanishing components with $l=2$: $A_{2,0,0}=\sqrt{\pi/5}$ with $\omega_{2,0,0}=0$, which only causes a static deformation, and $A_{2,2,2}=-\sqrt{6\pi/5}$ with $\omega_{2,2,2}=2(\Omega_o-\Omega_s)$, which is the asynchronous tide. Tidal forcing frequencies are often low such that $|\omega_{l,m,n}|\ll \omega_d$. Often only the quadrupolar terms with $l=2$ are retained (as in Eq.~\ref{Psi1}) and higher order terms in Eq.~\ref{Psi} are neglected, which is a reasonable approximation if $R_1/a\ll 1$. 

\subsection{Tidal response}

The tidal potential $\Psi$ produces forces that act upon body 1 to deform and excite internal flows within it. The goal of tidal theory is to determine the resulting response in body 1, representing its deformation and flows, and to compute the resulting effects on spins and orbits. Gravity is the way in which the two bodies interact (ignoring magnetic interactions etc.) so we must consider the gravitational potential perturbation $\Phi'$ of body 1, which can be evaluated outside body 1 to determine its effects on body 2. Since it satisfies $\nabla^2\Phi'=0$ outside body 1 (Poisson's equation evaluated in the zero density vacuum outside body 1 -- assumed spherical -- reduces to Laplace's equation) it can be expressed in the form \citep[e.g.~Section 21.3.1 of][]{RHB2006}
\begin{align}
\label{Phiexp}
\Phi'(\boldsymbol{x},t) =  \mathrm{Re}\left\{ \sum_{l=2}^{\infty}\sum_{m=0}^{l}\sum_{n=-\infty}^{\infty}\mathcal{B}_{l,m,n}\left(\frac{R_1}{r}\right)^{l+1} Y_l^m(\theta,\phi)\mathrm{e}^{-\mathrm{i}\omega_{l,m,n} t}\right\}.
\end{align}
The goal of tidal theory is to determine $\mathcal{B}_{l,m,n}$ for each component and hence the evolution of orbits and rotations. A useful way to quantify the response is in terms of the tidal Love number of each component
\begin{align}
k_{l,m,n} = \frac{\mathcal{B}_{l,m,n}}{\mathcal{A}_{l,m,n}},
\end{align}
the ratio of the response to forcing amplitudes, which is in general a dimensionless complex number and a function of tidal frequency and the orbital elements, where we have defined $\mathcal{A}_{l,m,n} = \epsilon_T \omega_d^2 A_{l,m,n}R_1^l/a^{l-2}$. The real part $\mathrm{Re}\{ k_{l,m,n} \}$ represents the in-phase component of the tidal deformation of the body. The imaginary part $\mathrm{Im}\{ k_{l,m,n} \}$ represents the out-of-phase component of the tidal response, which is responsible for transfers of energy and angular momentum between bodies 1 and 2 and hence is required to compute tidal evolution of orbits and rotations. If a component of the tide is perfectly in-phase with the corresponding component in $\Psi$, then $\mathrm{Im}\{ k_{l,m,n} \}=0$, and there is no tidal torque due to it and hence no corresponding angular momentum exchanges between bodies 1 and 2. For the $l=m=n=2$ tide, $\mathrm{Im}\{ k_{2,2,2} \}=0$ means that the ``tidal bulges" are instantaneously aligned with the position of body 2 as it orbits body 1, whereas a nonzero value of $\mathrm{Im}\{ k_{2,2,2} \}$ would imply a misalignment between the line of centres of the two bodies and the direction of the bulges, and hence there would be a resulting tidal torque (as a result of the dissipation of tidal energy causing the bulge to ``lag" behind the companion). Another popular way to express $\mathrm{Im}\{ k_{l,m,n} \}$ is in terms of the tidal quality factor \citep[e.g.][]{Goldreich1963}, an inverse measure of the dissipation,
\begin{align}
\label{Qdefn}
Q_{l,m,n} = \frac{2\pi E_{l,m,n}}{\int D_{l,m,n} \, dt},
\end{align}
which is proportional to the maximum tidal energy stored in each component $E_{l,m,n}$ to the energy dissipated in one (tidal) period $\int D_{l,m,n} \, dt$, where $D_{l,m,n}$ is the dissipation rate of that component. This is related to $\mathrm{Im}\{ k_{l,m,n} \}$ by
\begin{align}
\mathrm{Im}\{ k_{l,m,n} \} = \text{sgn}(\omega_{l,m,n})\frac{\mathrm{Re}\{ k_{l,m,n} \}}{Q_{l,m,n}} \equiv \text{sgn}(\omega_{l,m,n})\frac{k_{l}^{\text{h}}}{Q'_{l,m,n}},
\end{align}
which defines the modified tidal quality factor $Q'_{l,m,n}$, where $\text{sgn}(\omega_{l,m,n})=\pm 1$ indicates the sign of $\omega_{l,m,n}$. Here $k_{l}^{\text{h}}$ is the corresponding Love number of a homogeneous fluid body. $\mathrm{Re}\{ k_{l,m,n} \}$ is typically approximated by its hydrostatic value, $k_l$ ($k_2=0.0351$ for the Sun), and is typically a small number for centrally condensed stars. For quadrupolar tides with $l=m=2$, it can be shown that $k_{2}^{\text{h}}=3/2$ and hence if we omit subscripts on this component $Q'=\frac{3Q}{2k_2}$. We often use $Q'$ in preference to $Q$ to encapsulate our ignorance of $k_2$ (i.e.~internal structure) and because the combination $k_2/Q$ is what appears in tidal evolutionary equations (see below). 

It should be noted that there is no physical reason whatsoever for all components of the tide with different $(l,m,n)$ and tidal frequencies to have the same $Q'_{l,m,n}$ or $\mathrm{Im}\{ k_{l,m,n} \}$, even when considering only $l=2$ tides in linear theory. Simple tidal models such as the constant time-lag model of \citet{Darwin1880,Alexander1973,Mignard1980,Hut1981,EKH1998} assume that $Q'_{l,m,n}=\text{sgn}(\omega_{l,m,n}) k_l \omega_{l,m,n} \,\tau /k_l^{\text{h}}$, where $\tau$ is a constant (frequency-independent) lag time (for all components). This has an attractive simplicity and elegance in that evolutionary equations can be derived in closed form (and readily solved numerically) for all eccentricities and obliquities \citep[see also][and many others]{BO2009}, but this assumption is a very strong one, and it is not typically justified in stars\footnote{With the possible exception of tidal evolution involving giant stars if the dominant mechanism is a frequency-independent effective viscosity acting on equilibrium tides (see \S~\ref{EqmDamp}).}. This model is thus unlikely to be quantitatively correct in most scenarios and may not even be qualitatively correct to describe tidal evolution. Another popular simple model has $Q'_{l,m,n}$ the same and constant for every component, which can be related to a phase-lag angle that is the same for all tidal components. This is also usually unphysical \citep[see other objections in e.g.][]{EW2009,EM2013}.

\subsection{Simplest example of tidal spin-orbit evolution: circular (aligned) orbit with asynchronous spin}
\label{async}

Using only the definition in Eq.~\ref{Qdefn} (and that of $Q'$, restricted to $l=m=2$, and dropping subscripts for simplicity) and basic Keplerian orbital dynamics \citep[see e.g.][]{MD1999,Tremaine2023}, we can derive tidal evolutionary equations. Consider again asynchronous tides for a circular, aligned orbit ($e=i=0$), such that $\omega=2(\Omega_o-\Omega_s)$. The total (orbital + rotational) energy ($E$) and angular momentum ($L$), and their rates of change (indicated by overdots and obtained using the chain rule), are
\begin{align}
\label{EL}
E=\frac{-GM_1M_2}{2a} + \frac{1}{2}I_1\Omega_s^2, \quad L = \frac{M_1M_2}{(M_1+M_2)}\sqrt{G (M_1+M_2) a}+I_1\Omega_s, \\
\dot{E} = \frac{GM_1M_2}{2a} \frac{\dot{a}}{a}+I_1\Omega_s\dot{\Omega}_s, \quad \dot{L}=\frac{M_1M_2}{(M_1+M_2)}\sqrt{G (M_1+M_2) a}\frac{\dot{a}}{2a}+I_1\dot{\Omega}_s,
\label{ELdot}
\end{align}
where $I_1=r_{g}^2M_1 R_1^2$ is the moment of inertia of body 1 and $r_g^2$ is its squared radius of gyration\footnote{Defined by $r_g^2=\frac{8\pi}{3M_1R_1^2}\int_0^{R_1}\rho r^4\mathrm{d}r$, where $\rho(r)$ is the stellar density profile and $r$ is the spherical radius from the stellar centre.} ($r_g^2=0.0735$ for the Sun).
If the total angular momentum is conserved (neglecting interactions with other bodies, mass loss or magnetic braking of the stellar rotation), $\dot{L}=0$, so we may rewrite the tidal dissipation rate $D\equiv -\dot{E}$ by eliminating $\dot{\Omega}_s$ and rearranging to derive an evolutionary equation for the semi-major axis
\begin{eqnarray}
\label{adot}
\frac{\dot{a}}{a} =-2D\frac{M_1+M_2}{M_1M_2}\frac{1}{\Omega_o^2 a^2}\left(1-\frac{\Omega_s}{\Omega_o}\right)^{-1}=-6\,\text{sgn}(\omega)\frac{M_1+M_2}{M_1M_2}\frac{1}{\Omega_o a^2}\frac{E_0}{Q' k_2}=-\frac{9}{2}\text{sgn}(\omega)\frac{\Omega_o}{Q'}\left(\frac{M_2}{M_1}\right)\left(\frac{R_1}{a}\right)^5\propto \frac{-\text{sgn}(\omega)}{Q'a^{\frac{13}{2}}},
\end{eqnarray}
using the definition $D=\text{sgn}(\omega)\omega E_0/Q=3\text{sgn}(\omega)\omega E_0/(2Q'k_2)$ and noting that it can be shown that the peak tidal potential energy stored is\footnote{This comes from the peak potential energy stored \citep[\S 4.9 of][]{MD1999}, noting that the mass in the tidal bulges is approximately $\delta m\approx k_2 M_1 \xi_r/R_1$ \citep[e.g.][noting that his $k=k_2/2$]{Hut1981}. We obtain (crudely) $E_0\approx \mathrm{max}[|g \delta m_1 \xi_r|] \approx \mathrm{max}[|k_2M_1(\xi_r/R_1)\Psi |]\approx k_2M_1\epsilon_T\mathrm{max}[|\Psi|]\approx k_2M_1\epsilon_T^2\omega_d^2R_1^2A_{2,2,2}\mathrm{max}[|Y_2^2(\theta,\phi)\mathrm{e}^{-\mathrm{i}\omega t}|]= (3/4)k_2\frac{GM_1^2}{R_1}\epsilon_T^2\mathrm{max}[|\sin^2\theta\, \mathrm{e}^{2\mathrm{i}\phi-\mathrm{i}\omega t}|]$, where we have used $|\xi_r|\approx \Psi/g$ (neglecting $\Phi'$, see \S~\ref{EqmTide}), used Eq.~\ref{PsiAmp}, and the numerical factor comes from those in the $l=m=2$ tidal component, i.e. $\frac{1}{4}\sqrt{\frac{15}{2\pi}}\sqrt{\frac{6\pi}{5}}=\frac{3}{4}$. For an alternative, more rigorous, derivation of tidal torques see e.g.~\citet{Ogilvie2013}.} $E_0 = \frac{3}{4}k_2 \epsilon_T^2 GM_1^2/R_1$.  Conservation of angular momentum then implies that the rotation of body 1 evolves according to 
\begin{align}
\label{omdot}
\dot{\Omega}_s = -\frac{M_1M_2}{M_1+M_2} \frac{\Omega_o a^2}{2 I_1}\frac{\dot{a}}{a}=\frac{9}{4r_g^2}\text{sgn}(\omega)\frac{\Omega_o^2}{Q'}\frac{M_2}{M_1}\frac{M_2}{M_1+M_2}\left(\frac{R_1}{a}\right)^3 \propto \frac{\text{sgn}(\omega)}{Q' a^6}.
\end{align}
Since the overall sign of semi-major axis evolution is determined by $-\text{sgn}(\omega)=\text{sgn}(\Omega_s-\Omega_o)$, this tells us that orbits shrink ($a$ decreases) if $\Omega_o>\Omega_s$ ($P< P_s$ and $\Omega_s$ consequently increases; relevant for hot Jupiters orbiting slowly rotating stars) and expand ($a$ increases) if $\Omega_o<\Omega_s$ ($P> P_s$ and $\Omega_s$ consequently decreases; relevant for planets orbiting rapidly rotating stars, the Earth-Moon system, and Jupiter's \& Saturn's moon systems). A simple timescale to estimate how long it would take for significant rotational evolution of body 1 (e.g. towards synchronism) can be obtained using $\tau_{\Omega_s} \equiv |\Omega_s/\dot{\Omega}_s| \propto Q'$, and a similar estimate can be made for orbital evolution using $\tau_a = \frac{2}{13} |a/\dot{a}|\propto Q'$ (where the factor of $2/13$ comes from integrating in time $\dot{a}/a \propto a^{-13/2}$, assuming $Q'$ to be independent of $a$ and time). Timescales for tidal evolution of the orbit and rotation therefore depend on how efficiently tides are dissipated, and strongly on orbital separation $a$.

Tidal evolution of $\Omega_s$ or $a$ will not occur in this system if $\Omega_o=\Omega_s$ (i.e., $P=P_s$), which indicates a tidal equilibrium state, sometimes referred to as tidal spin-orbit synchronisation or ``tidal locking". The rotation and orbital periods are then equal so body 2 always sees the ``same side" of body 1 (at least in a planet with a solid surface on a circular, aligned orbit). Tidal dissipation inside the Moon's interior (treating the Moon as body 1 and the Earth as body 2) therefore explains why there is a ``far side of the Moon"\footnote{Not the title of a Pink Floyd album because the Moon does \textit{not} always show the same face to the Sun.} and why the Moon's axial rotation period matches its orbital period around the Earth. In close stellar binaries, tidal dissipation would also be expected to produce spin-orbit synchronisation (if $Q'$ is small enough), meaning that we would preferentially expect binaries with short enough orbital periods to have $P\approx P_s$. This is indeed what is found observationally, broadly speaking (see \S~\ref{Obs}).

Stars are known to lose mass and angular momentum by stellar winds \citep[as first theorised for the Sun by][]{Parker1958}, in which magnetic fields can enforce the material to co-rotate with the star out to ten or more stellar radii and substantially enhance the angular momentum loss \citep[e.g.][]{WD1967}. This is thought to explain why stars are observed to rotate more slowly as they age on the main sequence following the ``Skumanich law", $\Omega_s\propto t^{-\frac{1}{2}}$ \citep{Skumanich1972}, which can be modelled by introducing an additional term on the right hand side of Eq.~\ref{omdot} $\propto -\Omega_s^3$.

To illustrate qualitatively the tidal evolution possible for a two-body system with a circular orbit and asynchronous spin, we can transform Eqs.~\ref{adot} and \ref{omdot} by rescaling the variables and time to eliminate explicit dependence on $Q'$ -- which just determines the \textit{rate} of pure tidal evolution -- to obtain the following ordinary differential equations for the rescaled stellar rotation $\tilde{\Omega}_s=\Omega_s \sqrt{R_1^3/(G(M_1+M_2))}C^{-\frac{3}{4}}$ and orbital angular frequency $\tilde{\Omega}_o=\Omega_o \sqrt{R_1^3/(G(M_1+M_2))}C^{-\frac{3}{4}}\propto a^{-3/2}$, where $C=M_2/(r_g^2(M_1+M_2))$  \citep{Counselman1973,Hut1981,BO2009}. The derivation is left as an exercise, and the result (incorporating the magnetic braking term in {\color{blue} blue}) is:
\begin{align}
\label{rescaled1}
\dot{\tilde{\Omega}}_s=\tilde{\Omega}_o^4 \left(1-\frac{\tilde{\Omega}_s}{\tilde{\Omega}_o}\right) {\color{blue}{-A \tilde{\Omega}_s^3}}, \quad
\dot{\tilde{\Omega}}_o=3\tilde{\Omega}_o^{\frac{16}{3}}\left(1-\frac{\tilde{\Omega}_s}{\tilde{\Omega}_o}\right).
\end{align}
\begin{wrapfigure}{r}{0.45\textwidth}
\vspace{-1.2cm}
  \begin{center}
\includegraphics[width=0.45\textwidth,trim=1.5cm 1cm 4cm 1cm,clip=true]{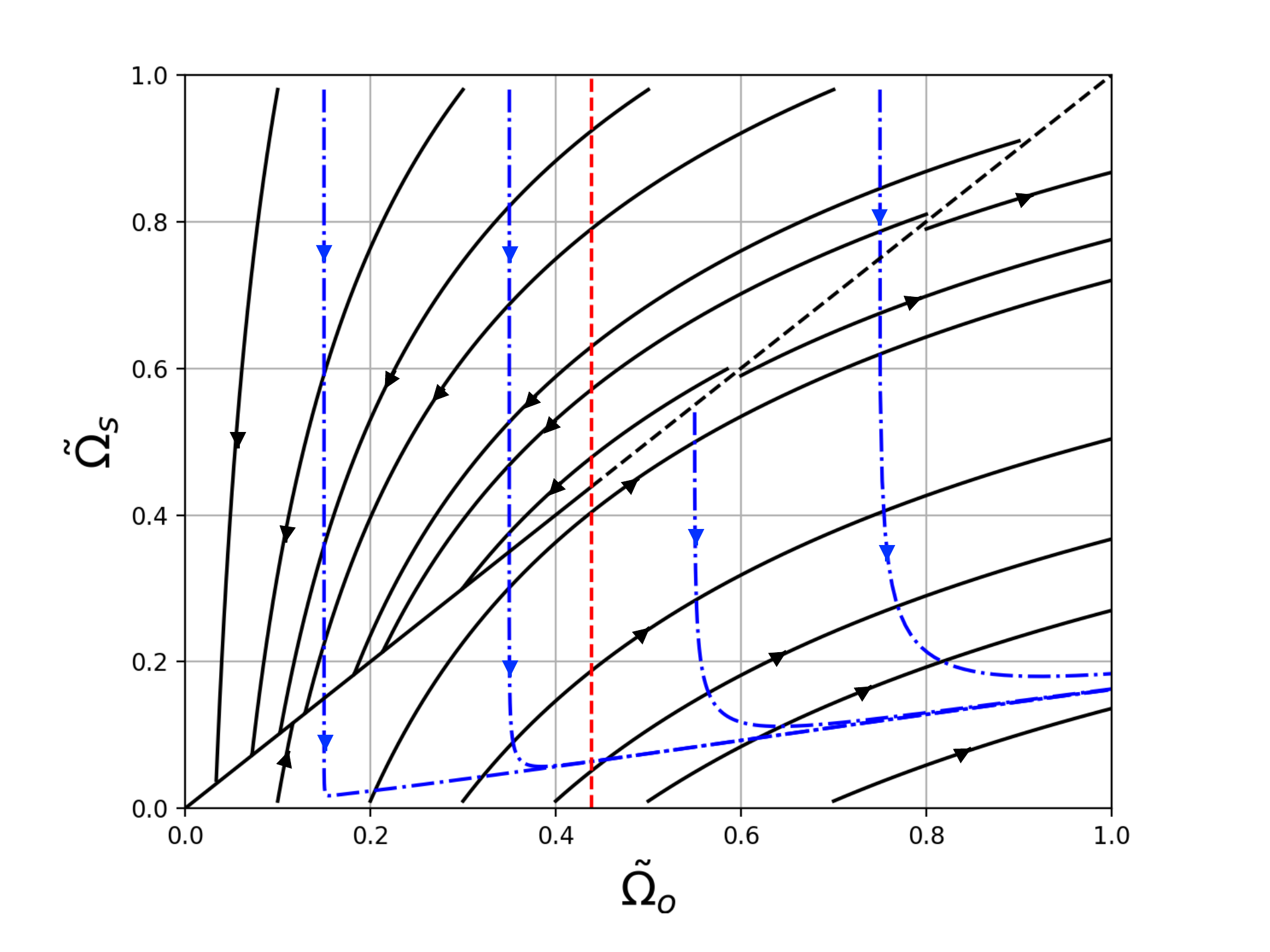}
\caption{Tidal evolution trajectories following the directions of arrows from the given initial conditions. Blue dot-dashed lines incorporate magnetic braking assuming $A=100$. Tidal equilibrium $\tilde{\Omega}_s=\tilde{\Omega}_o$ is stable (solid black line) for $\tilde{\Omega}_o\leq 3^{-\frac{3}{4}}$ (transition demarcated by red dashed line) and unstable otherwise (dashed black line). Trajectories going off to the right (if $\tilde{\Omega}_o\to \infty$) indicate the ultimate coalescence of the bodies.}
\label{chap1:fig2}
  \end{center}
\vspace{-1.2cm}
\end{wrapfigure}
This system (with $A=0$) has an equilibrium state corresponding to spin-orbit synchronisation where $\tilde{\Omega}_s=\tilde{\Omega}_o$, but this is unstable for $\tilde{\Omega}_o>3^{-\frac{3}{4}}\approx 0.439$ \citep[][]{Counselman1973,BThesis2011}, equivalent to the statement that no more than a quarter of the total angular momentum can be in the form of spin angular momentum for stability \citep{Hut1980}. Fig.~\ref{chap1:fig2} shows some representative solutions of Eqs.~\ref{rescaled1}, illustrating that orbits with $\tilde{\Omega}_s>\tilde{\Omega}_o$ generally expand ($\tilde{\Omega}_o$ decreases) until synchronism is reached, whereas orbits with $\tilde{\Omega}_s<\tilde{\Omega}_o$ generally shrink until either reaching synchronism (if there is enough angular momentum, where $L \propto \tilde{\Omega}_o^{-1/3}+\tilde{\Omega}_s$) or result in the inward migration (and, potentially, ultimate destruction) of body 2 into body 1 ($\tilde{\Omega}_o\to \infty$). The qualitative evolution is modified by other processes that affect stellar rotation; notably, magnetic braking that is thought to slow down stellar spins according to $\dot{\Omega}_s\propto- \Omega_s^3$, which has the effect of introducing the blue term in Eqs.~\ref{rescaled1}, and the new parameter $A$, which depends on $Q'$ so that the qualitative tidal evolution now depends on the rate of tidal evolution (whereas trajectories are universal, independent of $Q'$, if $A=0$). Some example solutions are shown in Fig.~\ref{chap1:fig2} using blue dot-dashed lines for $A=100$ \citep[see][for further details]{BO2009}, which show that the gradual angular momentum loss through magnetic braking prevents spin-orbit synchronisation from being a true equilibrium state. More sophisticated calculations of the existence and stability of tidal equilibrium states are studied in the elegant paper of \citet{Hut1980}. 

\subsection{Tidal spin-orbit evolution of a weakly eccentric orbit: orbital circularisation}
\label{weake}

Consider now tidal dissipation inside body 1 on an initially weakly eccentric orbit with $e\ll 1$, but neglect changes in stellar rotation and assume that the orbital angular momentum is constant (i.e.~assume stellar rotational angular momentum is negligible). It is known that a circular orbit is the one with the least energy for a given angular momentum, so we expect tidal dissipation to drive the orbit to become circular and hence reduce the eccentricity $e$. The semi-major axis of the circular orbit ($a_\circ$) with the same angular momentum ($\propto \sqrt{a(1-e^2)}$) as the eccentric orbit is thus related to that of the currently eccentric orbit by $a_\circ=a(1-e^2)$. Hence, the orbital energy (Eq.~\ref{EL}) of the eccentric orbit can be approximated as
\begin{align}
E=-\frac{GM_1M_2}{2 a} =\underbrace{-\frac{GM_1M_2}{2 a_\circ}}_{E_\circ} + \underbrace{\frac{GM_1M_2}{2 a_\circ} e^2}_{E_{\text{epi}}},
\end{align}
where $E_\circ$ is the final orbital energy of the equivalent circular orbit, and $E_{\text{epi}}$ is the remaining ``epicyclic orbital energy" which can be dissipated. Equating $-\dot{E}_{\text{epi}}=D$ and realising that $|\omega|=\Omega_o$ for eccentricity tides (neglecting any asynchronous rotation) provides us with
\begin{align}
\label{edot}
\frac{\dot{e}}{e} = -\frac{a_o D}{G M_1M_2 e^2} \propto -\frac{\Omega_o}{Q'}\left(\frac{M_2}{M_1}\right)\left(\frac{R_1}{a}\right)^5 \propto -\frac{1}{Q' a^{\frac{13}{2}}},
\end{align}
using the definition of $Q'$ for the relevant tidal component, since it can be shown that $E_0\propto e^2 \epsilon_T^2 k_2 GM_1^2/R_1$, ignoring $O(1)$ constants. Hence, an initially eccentric orbit will tend to circularise ($\dot{e}<0$ such that $e\to 0$ for sufficiently large times) due to tidal dissipation. This explains why the closest stellar binaries (for solar-type stars, approximately those with $P<10$ days) have been observed to have smaller eccentricities than those with longer periods, and tend to be circular for the shortest periods, at least if $Q'$ is small enough for this process to have occurred and if these systems start out eccentric. The picture is not quite so simple when the stellar rotation is allowed to change, where it is possible for large enough $\Omega_s$ to excite $e$ rather than damp it \citep[e.g.][]{Hut1981}. A crude timescale for significant evolution of the eccentricity (assuming all terms on the right hand side of Eq.~\ref{edot} are constant) can be obtained by $\tau_e=|e/\dot{e}| \propto Q'$. 

\subsection{Tidal equilibrium state}
\label{eqmstate}

If tidal dissipation and therefore spin-orbit evolution in a two-body system is efficient, or if tides can be assumed to have had an infinite time to act (ignoring other evolutionary processes for now), the final outcome will either be a tidal equilibrium state or the coalescence of the two bodies. Using an elegant and simple approach, \citet{Hut1980} has demonstrated that as long as there is sufficient angular momentum in a two-body system (where the two bodies rotate with, in general, different angular velocities $\Omega_{s,j}$ and obliquities $i_j$ for each of body $j=1$ and body $j=2$), there exists an equilibrium state in which
\begin{align}
\Omega_o = \Omega_{s,1} = \Omega_{s,2}, \quad e=0, \quad \textrm{and} \quad i_1=i_2=0.
\end{align}
This corresponds with a circular orbit for the two-body system, where each body has an aligned (i.e.~the equatorial planes of both bodies coincide with the orbital plane) and synchronous spin. This is the minimum energy state for a given angular momentum. We have illustrated special cases of this general result in \S~\ref{async} and \ref{weake}. It should be remembered that other processes, such as magnetic braking of stellar spins, or gravitational interactions with additional bodies, may prevent this equilibrium state from being reached in any given system. 

If there is insufficient angular momentum, or if less than three quarters of the total (orbital plus rotational) angular momentum is in the form of orbital angular momentum, the two bodies will eventually merge instead of reaching this equilibrium state. This is sometimes referred to as ``the Darwin instability" following\footnote{The second son of the famous naturalist Charles Darwin.} \citet{Darwin1879}, and is discussed in e.g.~\citet{Counselman1973,Hut1980}. It is just another name for the behaviour we have already discussed regarding Fig.~\ref{chap1:fig2} when the equilibrium state is unstable. The criterion mentioned there, $\tilde{\Omega}_o>3^{-\frac{3}{4}}$, can be re-expressed in the more familiar form
\begin{align}
a<R_1 \sqrt{\frac{3 r_g^2 (M_1+M_2)}{M_2}},
\label{Darwin}
\end{align}
such that sufficiently close orbits satisfying this condition are subject to tidally-driven orbital decay towards coalescence. For example, this criterion is consistent with observations of W UMa stellar binary systems, contact binaries for which there exists a lower limit on the mass ratio ($M_2/M_1$) below which none are observed. This is interpreted as the critical value below which systems are driven towards coalescence by tidal forces \citep[e.g.][]{Rasio1995}. In addition, the red luminous nova V1309 Sco probably resulted from a merger between two contact binary stars driven by this mechanism \citep[e.g.][]{Tylenda2011,Stepien2011}. In extrasolar planetary systems, sufficiently close-in planets may satisfy Eq.~\ref{Darwin} currently, in which case they would be expected to gradually spiral into their stars and be destroyed (though not necessarily within stellar lifetimes), or their stars may spin down sufficiently by magnetic braking so that they end up rotating more slowly than the orbital periods of their planets, so a stable equilibrium state cannot be reached \citep[e.g.][]{BO2009,Soko2010,Damiani2015}. We will revisit observational evidence for planetary orbital decay and destruction in \S~\ref{Obs}.

\subsection{Summary and further topics}

We have introduced the basic aspects of tidal evolution, and highlighted why we must understand the tidal responses of stars, and particularly determine their tidal dissipation rates $D_{l,m,n}$ (or Love numbers $k_{l,m,n}$ or quality factors $Q'_{l,m,n}$) to model the evolution of many stellar and planetary systems. In the next section we will explore the tidal response of a fluid body. This is where the interesting fluid dynamics (or magnetohydrodynamics) comes in.

\section{Tidal flows in stars and planets} \label{response}

The fluid response of a star or gaseous planet to its companion's tidal perturbation is often decomposed into two components: an equilibrium or non-wavelike tide, and a dynamical or wavelike tide. For small tidal amplitudes ($\epsilon_T\ll 1$) where linear theory is appropriate, the tidal response can be split up formally in any such way that is convenient as long as the governing equations and boundary conditions are satisfied. For larger amplitude tides where nonlinear effects are important, the decomposition is less clear cut but may still be beneficial for interpretation.

\subsection{Equilibrium (non-wavelike) tides}\label{EqmTide}

These are the instantaneous quasi-hydrostatic and adiabatic (neglecting dissipation, heat sources/sinks) deformation of the body, and the associated flow inside it that moves the tidal bulges around to follow the companion. It is what most people think of as ``the tide", but in many problems it is unlikely to provide the most important contribution to \textit{dissipation} and hence spin-orbit evolution, in a similar way that the ``barotropic tide" (the name geophysicists use) doesn't dominate tidal dissipation in Earth's oceans. For illustration, consider a slowly-rotating\footnote{So we can neglect centrifugal forces and approximate the body as spherical, which is valid if $\Omega_s^2/\omega_d^2\ll 1$.} spherically-symmetric star in hydrostatic equilibrium, such that
\begin{equation}
\label{hydrostatic}
    \nabla p=\rho\boldsymbol{g}=-\rho\nabla\Phi,
\end{equation}
where $p(r)$ is the pressure, $\rho(r)$ is the density, $\boldsymbol{g}=-g(r)\boldsymbol{e}_r=-\nabla \Phi$ is the gravitational acceleration, and $\Phi(r)$ is the gravitational potential satisfying Poisson's equation $\nabla^2\Phi = 4\pi G\rho$, where we again adopt spherical polar coordinates $(r,\theta,\phi)$ centred on the star. The buoyancy, or Brunt-V\"{a}is\"{a}l\"{a}, frequency, $N(r)$, measures the frequency at which a fluid parcel will oscillate vertically if it is perturbed in a stably stratified (i.e.~stable to convection) atmosphere, and is defined by\footnote{See section 3.3.2 of the excellent lecture notes here: \url{https://users-phys.au.dk/jcd/oscilnotes/Lecture_Notes_on_Stellar_Oscillations.pdf}.}
\begin{equation}
    N^2 = g\frac{\mathrm{d}}{\mathrm{d} r} \ln\left(\frac{ p^{\frac{1}{\Gamma_1}}}{\rho}\right)= g\left(\frac{1}{\Gamma_1 p}\frac{\mathrm{d} p}{\mathrm{d} r}-\frac{1}{\rho}\frac{\mathrm{d} \rho}{\mathrm{d} r}\right),
\end{equation}
where $\Gamma_1=(\partial \ln p/\partial \ln \rho)_s$, and the subscript $s$ refers to constant specific entropy. Wherever $N^2<0$, this portion of the star is convectively unstable, meaning that when fluid parcels are perturbed upwards they continue to rise. Convective turbulence is usually assumed to be efficient in transporting heat, such that it approximately homogenises the entropy so that $N^2\approx 0$ (``convection eliminates the unstable gradients that drive it"). Stars can be fully convective (for M-type stars with masses below $0.4M_\odot$, with $M_\odot=2\times 10^{30} \mathrm{kg}$ the solar mass) or contain a mixture of radiative ($N^2>0$) and convective ($N^2\approx 0$) regions (cores and/or envelopes) for stars larger than approximately $0.4M_\odot$, so we must consider tides in both types of region. Solar type stars have convective envelopes and radiative cores. Stars more massive than $1.1M_\odot$ have convective cores, overlying radiative regions and also (depending on the stellar mass) thin convective envelopes.

\begin{wrapfigure}{r}{0.5\textwidth}
\vspace{-1.2cm}
  \begin{center}
\includegraphics[width=0.5\textwidth,trim=0cm 0cm 0cm 0cm,clip=true]{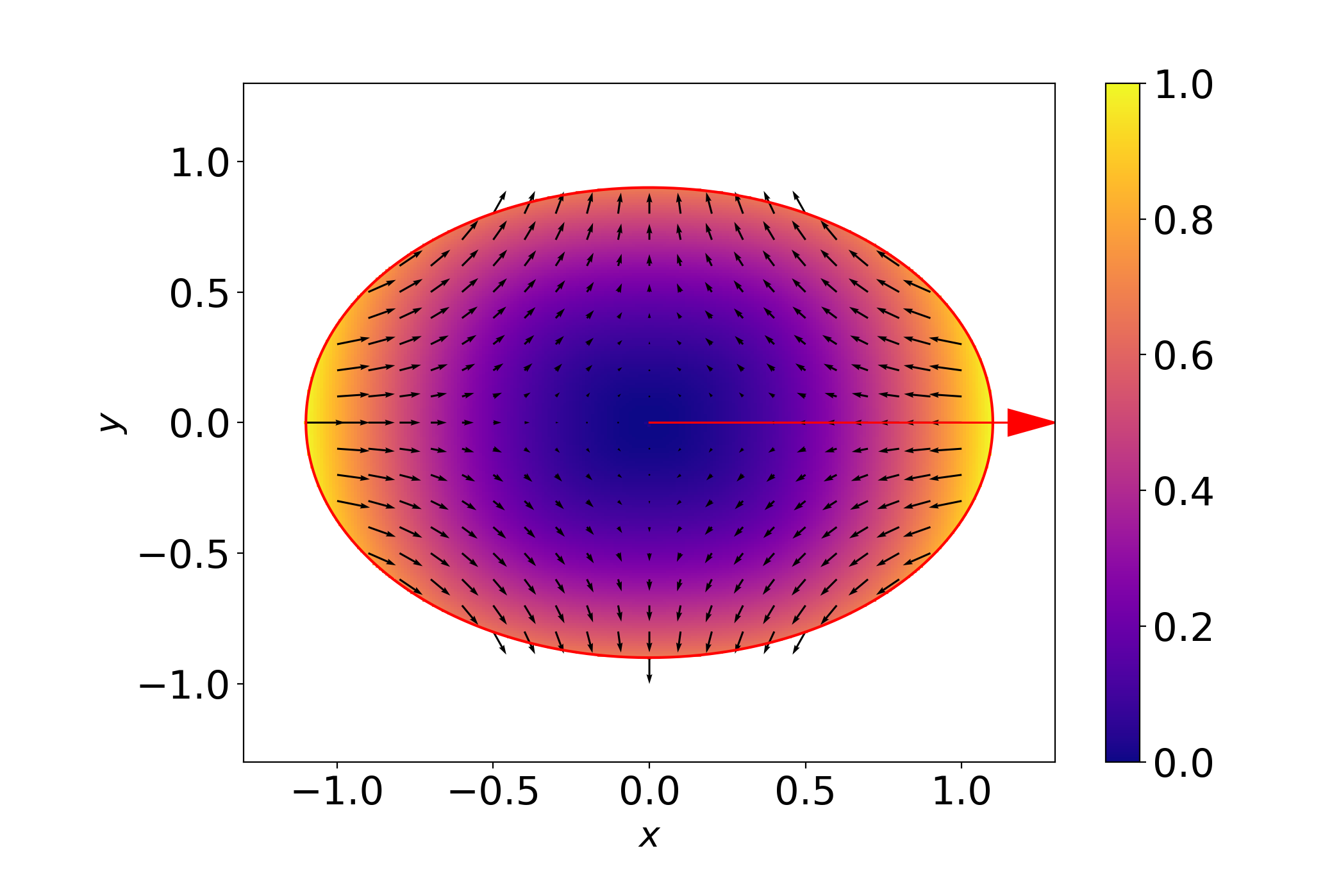}
\caption{Equilibrium tide velocity field (with $l=m=2$) in the stellar equatorial plane in the frame rotating with the star (here modelled as an $n=1$ polytrope). The companion is located along the $x$-axis at the chosen time, as indicated by the red arrow. Colours indicate velocity magnitude $|\boldsymbol{u}_e|$ (normalised to 1) and black arrows indicate local velocity vectors, which act to move around the tidal bulges to follow the companion. The boundary of the star is indicated as ellipsoidal with an ellipticity of 0.1 for visualisation purposes.}
\label{chap1:fig3}
  \end{center}
\vspace{-3cm}
\end{wrapfigure}
We perturb the star with the tidal potential $\Psi$. The equilibrium tide is computed assuming the body remains in instantaneous hydrostatic equilibrium, meaning we neglect inertial terms in the fluid momentum equation, which is appropriate if $\omega^2\ll \omega_d^2$ and $\omega^2\ll |N^2|$ (hence buoyancy forces can be neglected). It thus satisfies the momentum (Euler) equation
\begin{align}
\nonumber
\boldsymbol{0} &=-\frac{1}{\rho+\rho'}\nabla (p+p')-\nabla (\Phi+\Phi') -\nabla \Psi\\
&=-\frac{1}{\rho}\nabla p'+\frac{\rho'}{\rho^2}\nabla p - \nabla \Phi' - \nabla \Psi +\underbrace{O(\rho'^2,p'^2)}_{\text{neglect}},
\end{align}
where we have Taylor expanded density ($\rho+\rho'$), pressure ($p+p'$) and gravitational potential ($\Phi+\Phi'$) about the reference state $\rho,p,\Phi$, and used Eq.~\ref{hydrostatic}. Manipulating this allows us to derive equations for (Eulerian) perturbations to pressure, density and gravitational potential \citep[e.g.][]{Ogilvie2014}
\begin{align}
 p'&=-\rho(\Phi'+\Psi), \\
 \rho'&= -\frac{\mathrm{d}\rho}{\mathrm{d}p}\rho(\Phi'+\Psi),  \\
    \nabla^2\Phi' &= -4\pi G\frac{\mathrm{d}\rho}{\mathrm{d}p}\rho(\Phi'+\Psi),
\end{align}
inside the star, and $\nabla^2\Phi' = 0$ outside, assuming a vacuum there. After expanding $\Phi'$ using Eq.~\ref{Phiexp} and writing the coefficient of $Y_l^m \mathrm{e}^{-\mathrm{i}\omega t}$ as $\Phi_l$, and similarly for $\Psi$ using $\Psi_l$, the latter 
can be written for each $l$ as
\begin{eqnarray}
\label{PhiEqn}
\frac{1}{r^2}\frac{\mathrm{d}}{\mathrm{d}r}\left(r^2 \frac{\mathrm{d} \Phi'_l }{\mathrm{d}r}\right)-\frac{l(l+1)}{r^2}\Phi'_l+4\pi G\frac{\mathrm{d}\rho}{\mathrm{d}p} \rho(\Phi'_l+\Psi_l)=0.
\end{eqnarray}
This is a second-order differential equation in $r$ for each harmonic component, so $\Phi'_l$ must satisfy the two boundary conditions required for the solution to not diverge as $r\to 0$ or $r\to \infty$ \citep[e.g.~Section 21.3.1 of][]{RHB2006}
\begin{eqnarray}
\frac{\mathrm{d}\ln \Phi'_l}{\mathrm{d} \ln r} =\frac{r}{\Phi'_l}\frac{\mathrm{d} \Phi'_l}{\mathrm{d} r}= l \quad\text{at} \quad r=0, \quad\quad \frac{\mathrm{d}\ln \Phi'_l}{\mathrm{d} \ln r} =\frac{r}{\Phi'_l}\frac{\mathrm{d} \Phi'_l}{\mathrm{d} r}= -(l+1) \quad\text{at} \quad r=R_1.
\end{eqnarray}
The equilibrium tidal flow ($\boldsymbol{u}_e=\partial \boldsymbol{\xi}_e/\partial t$) can be defined by the displacement field $\boldsymbol{\xi}_e=\xi_{e,r}\boldsymbol{e}_r+\boldsymbol{\xi}_{e,h}$, where $\boldsymbol{\xi}_{e,h}\cdot \boldsymbol{e}_r=0$ . In radiative regions with $N^2>0$, this satisfies
\begin{equation}
\label{EQMtide}
    \xi_{e,r}=-\frac{\Phi'+\Psi}{g}, \quad\text{and}\text\quad \nabla\cdot\boldsymbol{\xi}_e=0,
\end{equation}
which is the conventional incompressible equilibrium tide \citep{Zahn1966,Zahn1989,Remus2012}. This solution does not apply in convective regions if they are well-mixed with $N^2\approx 0$ (more precisely when $\omega^2\nll |N^2|$), however, and we must instead compute the displacement in a different manner \citep{T1998,GD1998}. If $N^2=0$ (i.e.~perfectly mixed entropy), the equilibrium tide is irrotational, i.e.~$\nabla \times \boldsymbol{\xi}_e=\boldsymbol{0}$, so we can write  $\boldsymbol{\xi}_\mathrm{e}=\nabla X$, and $X$ is determined by the solution of \citep{T1998,Ogilvie2013}
\begin{eqnarray}
\label{NWLtide}
\nabla \cdot \left(\rho \nabla X\right)=\frac{\mathrm{d}\rho}{\mathrm{d}p}\rho \left(\Phi'+\Psi\right).
\end{eqnarray}
Expanding $X$ in terms of spherical harmonics as with $\Psi$ above gives, for each $l$,
\begin{eqnarray}
\label{Xeqn}
\frac{1}{r^2}\frac{\mathrm{d}}{\mathrm{d}r}\left(r^2\rho \frac{\mathrm{d} X_l }{\mathrm{d}r}\right)-\frac{l(l+1)}{r^2}\rho X_l=\rho\frac{\mathrm{d}\rho}{\mathrm{d}p}(\Phi'_l+\Psi_l).
\end{eqnarray}
This is also a second-order differential equation so requires two boundary conditions in each convective region: at the centre of a star (if it is convective there; or at a solid core) we have $\xi_{\mathrm{e},r}=\frac{\mathrm{d}X_l}{\mathrm{d}r}=0$ at $r=0$, and for all other boundaries, $\xi_{\mathrm{e},r}=\frac{\mathrm{d}X_l}{\mathrm{d}r}=-\frac{\Phi'+\Psi}{g}$, which also applies at the stellar radius $r=R_1$ for a convective envelope. The two formulations of equilibrium tide differ in general, though the distinction is not drastic in most cases. [We have ignored non-adiabatic effects, which may be important near the stellar surface \citep[e.g.][]{Bunting2019}.]

Once we have $\boldsymbol{\xi}_e$ from solving either Eqs.~\ref{EQMtide} or \ref{Xeqn}, the tidal velocity field is given by (the real part of) $\boldsymbol{u}_e = \partial_t \boldsymbol{\xi}_e=-\mathrm{i}\omega \boldsymbol{\xi}_e$ for each component with a given tidal frequency $\omega$. In linear theory we compute the response of each $l,m,\omega$ component independently and sum up the contributions to obtain the total flow. In Fig.~\ref{chap1:fig3} we illustrate the equilibrium tidal flow velocity field with $l=m=2$ in the equatorial plane of a star with a companion on a circular orbit at a fixed time when the companion lies along the $x$-axis (along the red arrow), using arrows to indicate the local velocity field at various points within the star, and colours to denote the (unit-normalised) magnitude of the velocity $|\boldsymbol{u}_e|$. The boundaries of the star are defined by the red ellipse $x^2/a^2+y^2/b^2=1$ with semi-axes $a=1.1$ and $b=0.9$ to show a strongly deformed star with radius 1 for visualisation purposes. Together with the bulk rotation of the star, this equilibrium tide makes the fluid streamlines elliptical in the reference frame rotating with the tidal bulges.

\subsection{Dynamical (wavelike) tides}\label{DynTide}

\begin{figure}[b]
\vspace{-0.5cm}
\begin{center}
\includegraphics[width=0.8\textwidth,trim=1cm 13cm 0.7cm 0.5cm,clip=true]{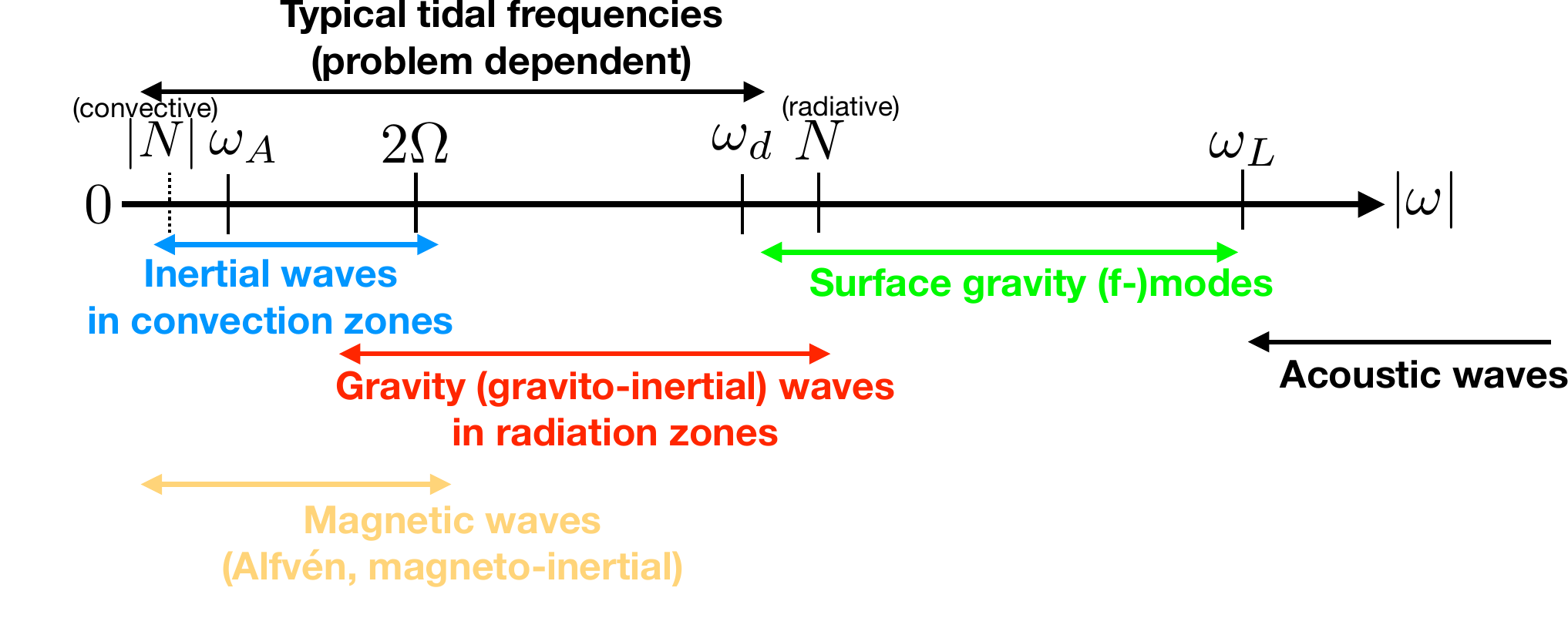}
\caption{Wave types in a rotating, magnetised star and their associated frequencies. ($|N|$ here applies to convective zones only.)}
\label{chap1:fig4}
  \end{center}
\vspace{-1cm}
\end{figure}
These consist primarily of waves of various kinds excited by tidal forcing. Stars exhibit waves with a variety of restoring forces (see Fig.~\ref{chap1:fig4}): 

\begin{itemize}
\item Acoustic waves (p modes or sound waves) restored by compressibility (gas pressure). These largely tend not to be excited resonantly by tidal forcing because tidal frequencies are typically much smaller than acoustic frequencies (which are bounded below for each $l$ by the Lamb frequency $\omega_L=\sqrt{l(l+1)}c_s/r$, where $c_s$ is the sound speed), but they can be excited in some triple systems \citep{Fuller2013} or in particularly violent tidal encounters \citep{PT1977}.
\item Surface gravity waves (f-modes or fundamental modes) restored by gravity/buoyancy forces at the surface of a star (or interfacial waves at an internal interface with a jump in density). These correspond with oscillatory deformations of the surface of a body that decay away from the surface. They tend not to be excited resonantly by tidal forcing for small eccentricities because tidal frequencies are typically much smaller than those of these waves, which are larger than but comparable to the dynamical frequency. Exceptions occur, particularly for highly eccentric orbits or tidal capture events though, where these may provide the dominant contribution \citep{PT1977,Lai1997,IvPap2007}. For highly eccentric orbits, chaotic amplification of f-modes to nonlinear amplitudes may occur and very efficiently circularise orbits \citep{Mardling1995,IvPap2004,Wu2018,Vick2018,Yu2022}, though the nonlinear behaviour of these waves is not well understood theoretically and further progress will likely require sophisticated direct numerical simulations.
\item Internal gravity waves (g-modes or gravity waves\footnote{Not to be confused with gravitational waves (``ripples in spacetime") in General Relativity, which have an entirely different nature.}) restored by buoyancy forces in stably-stratified (non-convective) radiative layers. Referred to as inertia-gravity or gravito-inertial waves when restored by a combination of buoyancy and Coriolis forces. These waves are excited in radiative zones and are thought to be particularly important in stars with outer radiative envelopes (early-type stars), but they can also be important in solar-type stars with radiative cores. These waves are best illustrated using the dispersion relation describing short-wavelength incompressible waves and ignoring the effects of boundaries. In spherical geometry, such waves (with solutions for pressure perturbations $\propto Y_l^m(\theta,\phi) \mathrm{e}^{-\mathrm{i}\omega t}$) propagate with frequencies satisfying 
\begin{align}
\omega^2= \frac{N^2 k_\perp^2}{k^2},
\end{align}
where $k_\perp=\sqrt{l(l+1)}/r$ is the horizontal wavenumber (depending on spherical harmonic degree $l$) and $k=\sqrt{k_r^2+k_\perp^2}$ is the total wavenumber (with $k_r$ the radial component, with corresponding wavelength $2\pi/k_r$). These satisfy $|\omega |\leq N$, but this condition is readily achieved in stellar radiation zones where $N\sim \omega_d$ since the forcing frequencies are typically low with $|\omega|\ll \omega_d$.
\item Inertial waves (Coriolis modes, of which r-modes are a special subset) restored by Coriolis forces in rotating bodies. In a body uniformly rotating at the rate $\boldsymbol{\Omega}_s$, short-wavelength inertial waves with wavevectors $\boldsymbol{k}$ propagate with frequencies satisfying the local dispersion relation
\begin{align}
\omega^2= \frac{(2\boldsymbol{k}\cdot\boldsymbol{\Omega}_s)^2}{k^2}.
\end{align}
These therefore satisfy $|\omega |\leq 2\Omega_s$ and are only excited by such (sufficiently low) tidal frequencies. This condition is readily satisfied in many tidal problems however, indicating that these waves are likely to be excited by tidal forcing in many scenarios.
\item Magnetic waves, including Alfv\'{e}n waves restored by magnetic tension in magnetised bodies (which have typical frequencies satisfying $\omega^2_A=(\boldsymbol{k}\cdot \boldsymbol{B})^2/(\mu_0\rho)$, where $\boldsymbol{B}$ is the magnetic field and $\mu_0$ is the vacuum permeability), as well as a variety of mixed waves e.g. magneto-inertial or magneto-Coriolis \citep{LO2018}, magneto-gravity \citep{DdVLB2024} and magneto-acoustic waves restored by a combination of magnetic tension and pressure. Magnetic effects have been very poorly explored to date; studying them further is an exciting avenue for future work.
\end{itemize}

\section{Mechanisms of tidal dissipation} \label{mechanisms}

We split these up into mechanisms primarily operating on either equilibrium or dynamical tides, though it should be noted that there is no perfect distinction between these. This is because some mechanisms acting on equilibrium tides excite waves, and some act on both components. We will discuss mechanisms operating in convection and radiation zones in turn. 

Since the Sun, and presumably any star, possesses internal differential rotation (driven by other processes, such as convection, ultimately powered by internal nuclear reactions, or magnetic braking spin-down torques acting on the outer layers), this would imply some ongoing tidal dissipation in differentially-rotating stars. [It is also possible in principle for tidal ``anti-dissipation" to occur \citep{OL2012,DBJ2020a,Fuller2021} if another energy source is present e.g.~due to convection, internal fluid instabilities or heat sources.]

\subsection{Convection zones}

Convection zones are regions where heat is transported by, usually highly turbulent, fluid motions with $N^2<0$. Buoyancy forces are thus destabilising, so gravity waves are not supported and are evanescent in convective regions (i.e.~they decay away with distance from neighbouring radiative regions). If convection zones are well mixed, they have $N^2\approx 0$ (but slightly negative to drive convection), so inertial (and magnetic) waves are typically the ones supported which are relevant for the tidal response.

\subsubsection{Equilibrium tides interacting with turbulent convection} \label{EqmDamp}
This is perhaps the most controversial mechanism though much progress has been made exploring it in recent years. It could operate in any convective region of a star and has been proposed in the past to be the dominant tidal mechanism in all stars with convective envelopes \citep[e.g.][]{Zahn1977}, though this is no longer a widely accepted view. There are many uncertainties remaining though, ultimately because this mechanism requires us to understand aspects of convective turbulence and of its interaction with tidal flows.

The basic idea often considered is that turbulent convective fluid motions can extract energy from large-scale tidal flows by acting as an effective (or turbulent) viscosity, which is generally much larger than the negligibly small molecular viscosity \citep{Zahn1966,Zahn1977}. This interaction may be crudely modelled using mixing-length theory in an analogous way to stellar evolution codes modelling the turbulent transport of heat by convection; the primary differences are that velocity gradients replace temperature (or entropy) gradients and we are considering momentum (and kinetic energy) rather than heat transport. This conventional phenomenological approach assumes convection at a given radial location in a star has typical velocity $u_c$, and length-scale\footnote{Usually thought of as the mixing length at which fluid elements transport momentum or heat, or the dominant scale of the convective eddies, which is typically a multiple (``the mixing-length parameter") of the pressure scale height, though it should be remembered that there is in reality a spectrum of convective eddies spanning a range of length scales.} $l_c$, then this effective viscosity has magnitude $\nu_E\propto u_c l_c$, at least on dimensional grounds (though of course any dimensionless function could multiply this factor and the proportionality constant is not determined this way). For simplicity, it is often assumed that the viscosity is isotropic (a scalar) and depends only on $r$, in which case the resulting tidal dissipation can be readily computed by integrating the local viscous dissipation at each radius throughout the convection zone(s) using
\begin{align}
D\approx \frac{1}{2}\int_{\mathrm{CZ}}\rho \nu_E \,||(\partial_i u_{e,j}+\partial_j u_{e,i})||^2 \mathrm{d}V \propto Q'^{-1},
\label{Deqm}
\end{align}
where $\mathrm{d}V$ is the spherical polar volume element, $\partial_i$ are components of the gradient operator and $u_{e,i}$ are those of the tidal velocity (and $||E||^2=E_{ij}E_{ij}$, the tensor contraction of the rank-2 tensor $E$, using the summation convention). Under the above assumptions, specifying a form for $\nu_E(r)$, and given a stellar model with a prescribed internal structure (such as $\rho(r)$), $D$ can be computed and hence so can $Q'$ and the resulting tidal spin-orbit evolution. Calculations indicate that $Q'\gtrsim 10^7$ for most tidal frequencies under the above assumptions in models of solar-like main-sequence (MS) stars \citep[e.g.~Fig.~3 of][]{B20}.
 
One important aspect, realised by \citet{Zahn1966}, is that this mechanism is likely to be substantially less effective for ``fast tides" that are faster than the convection \citep[see also][]{Zahn1989}. More specifically, if the dominant convective eddies have a turnover frequency $\omega_c\sim u_c/l_c$, ``fast tides" occur when tidal frequencies satisfy $\omega\gg \omega_c$, and in this regime $\nu_E$ should be reduced in some fashion, with \citet{Zahn1966} arguing specifically for a multiplicative factor of $\omega_c/\omega$. \citet{GN1977} proposed a different suppression of the turbulent viscosity for fast tides, arguing that resonant eddies (those in a turbulent cascade with turnover frequencies matching the tidal frequency) dominate the interaction with the tidal flow, and obtained $\nu_E\propto u_c l_c (\omega_c/\omega)^2$ instead. This controversy remained in the literature for decades \citep[see also][]{GO1997} 
until direct numerical simulations of the interaction between tidal and convective flows started to be performed \citep[e.g.][]{Penev2009,OL2012,DBJ2020b,VB2020a}.\begin{wrapfigure}{r}{0.6\textwidth}
\vspace{-0.5cm}
\includegraphics[width=0.6\textwidth,trim=0cm 0cm 0cm 0cm,clip=true]{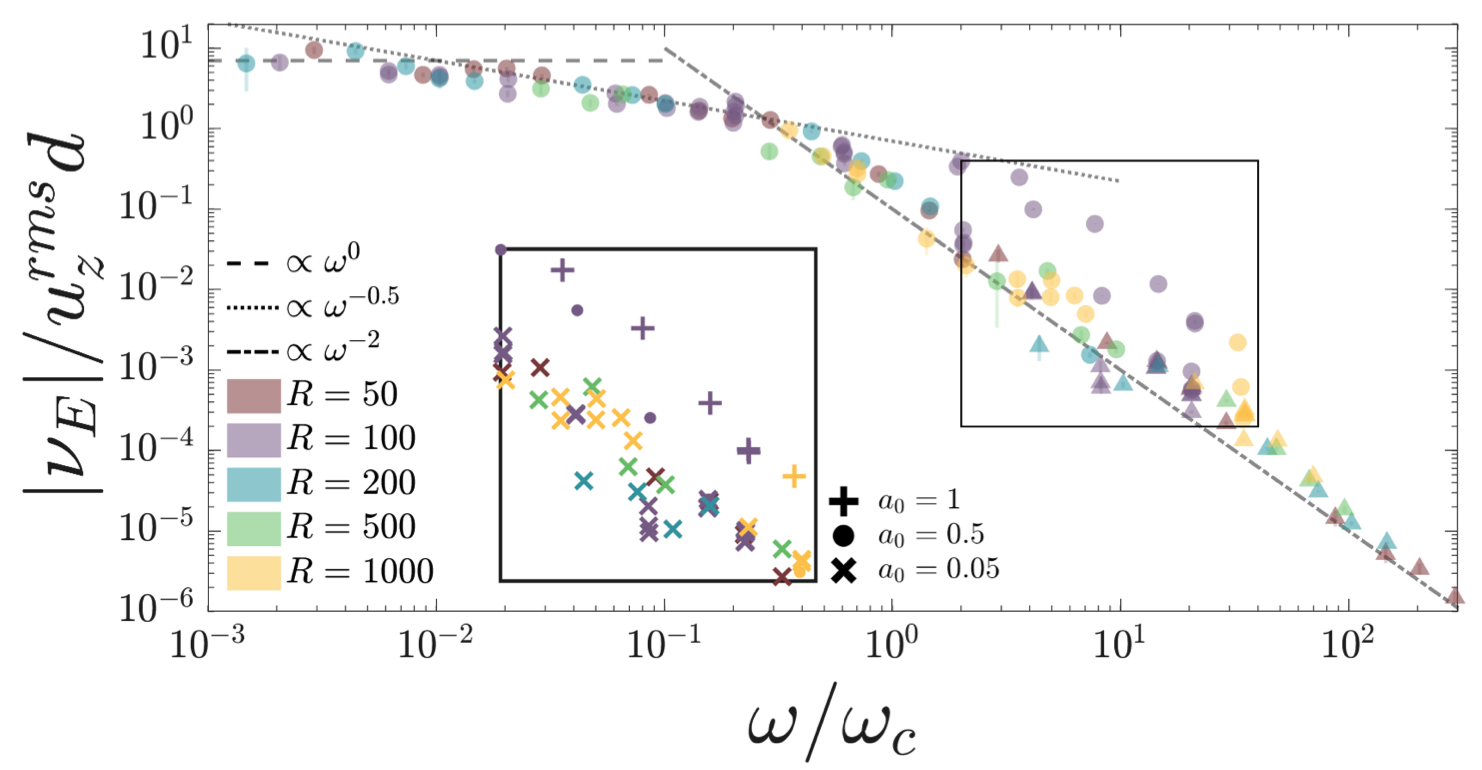}
 \caption{Effective (turbulent) viscosity (normalised by a measure of $u_c l_c$) as a function of the ratio of tidal to convective frequencies as obtained from direct numerical simulations of oscillatory tidal flows interacting with turbulent convection in a local ``small-patch" model of a portion of a convective zone of a star \citep[from Fig.~5 in][]{DBJ2020b}. ($R$ indicates how strongly the convection is driven and $a_o$ is related to the tidal amplitude.) For very fast tides, $\nu_E<0$ indicating tidal anti-dissipation (triangles).}
\label{nuE}
\vspace{-0.5cm}
\end{wrapfigure}
The current consensus from numerical simulations -- albeit ones using simplified models -- is that $\nu_E\propto (\omega_c/\omega)^2$ for $\omega\gg \omega_c$ -- though the reasons for this occurring do not appear to be compatible with the arguments of \citet{GN1977}, as shown by \citet{DBJ2020b} -- and that $\nu_E$ depends on tidal frequency except for very low frequencies $\omega/\omega_c\lesssim 10^{-2}$ (see Fig.~\ref{nuE}). The modern viewpoint, confirmed by asymptotic analysis for simple flows, is that convective flows respond viscoelastically to rapid oscillatory tidal flows, with a dominant elastic component, and a weaker viscous one \citep{OL2012,BravinerThesis2015,DBJ2020a} that falls off as $\omega^{-2}$ for very fast tides. [The incorporation of stellar/planetary rotation indicates that it can inhibit convective fluid motions and reduce the sizes of the dominant convective eddies, both of which reduce $\nu_E$ further for rapid rotation; \citealt{Mathis2016,dVBH2023}.]
When this frequency-reduction of $\nu_E$, consistent with numerical simulations, is accounted for in Eq.~\ref{Deqm}, the resulting tidal dissipation is typically substantially reduced in main-sequence stars (for example in hot Jupiter host stars, we expect $\omega/\omega_c\gtrsim 20$ for eddies in deeper convective layers), implying that this mechanism is not generally very effective. The resulting $Q'\gtrsim 10^{10}-10^{11}$ for most tidal frequencies in solar-like main-sequence stars \citep[e.g.~Fig.~3 of][]{B20}, leading to negligible tidal evolution during the stellar lifetime, such that, if existing prescriptions are at least approximately correct, this mechanism can be ignored in most main-sequence stars. There is much that remains to be explored further however.

More positively, this mechanism is probably the dominant one in evolved giant stars, which have much larger stellar radii, and where tidal frequencies are generally not so fast relative to convective frequencies for the drop-off of $\nu_E$ for fast tides to be crucial. As we will discuss further in \S~\ref{Obs}, this mechanism has been empirically confirmed against observations for these stars by \citet{VP1995} \citep[see also][]{Hansen2012,PW2018,Beck2018}. This mechanism is likely to play a key role in determining the ultimate fate of the Earth when the Sun becomes a red giant \citep[e.g.][]{Rasio1996,Mustill2012}. This mechanism has also been proposed to explain orbital circularisation of solar-type pre-main-sequence (PMS) stars (which have not yet begun burning hydrogen in their cores) by \citet{ZahnBouchet1989}, who also realised the importance of coupling calculations of tidal evolution with stellar structural evolution, particularly during the PMS phase. However, particularly if $\nu_E\propto \omega^{-2}$ for fast tides, this mechanism is incapable of explaining the circularisation of solar-type stars.

Recently, an alternative viewpoint to model the interaction between tidal flows and convection has also been proposed, specifically motivated by applications to fast tides. The above discussion assumes energy exchanges between tidal and convective flows occur primarily due to velocity gradients in the tidal flow (and arise from part of the nonlinear inertial term in the fluid momentum equation for the tidal $\boldsymbol{u}_e$ plus convective $\boldsymbol{u}_c$ flow of the form $\rho (\boldsymbol{u}_c\cdot \nabla) \boldsymbol{u}_e$, contributing to the rate of change of energy a term (before spatial integration) $\mathcal{I}_1=\rho \boldsymbol{u}_c\cdot (\boldsymbol{u}_c\cdot \nabla) \boldsymbol{u}_e$). Since this is small when integrated in time over a tidal period for fast tides, specifically it vanishes when $\omega/\omega_c\to \infty$ because $\nu_E\to 0$ there (and the interaction term is linear in the tidal flow), what about other possible interaction terms in the momentum equation? \citet{T2021} realised there is also an interaction term involving gradients of the convective flow and products of tidal flow components, with energy exchanges of the form (before spatial integration) $\mathcal{I}_2=\rho \boldsymbol{u}_e\cdot (\boldsymbol{u}_e\cdot \nabla) \boldsymbol{u}_c$), which should not vanish on time-integration as $\omega/\omega_c\to \infty$ (because the interaction term is not linear in the tidal flow). This has an apparent advantage in terms of its relative value (to $\mathcal{I}_1$) after time integration for very fast tides \citep[and its crudest evaluation -- assuming it to be positive everywhere before spatial integration -- appears to match observations of solar-type star orbital circularisation periods,][]{TM2021}. However, partly because convection typically occurs on much shorter length-scales than the equilibrium tidal flow, and because this term simply cannot be positive at every spatial location, there will be significant cancellations upon spatial integration (more so than expected for $\mathcal{I}_1$). \textit{It is the net contribution after both spatial and time integration (i.e.~over the entire star and over many tidal and convective periods) that is relevant for driving tidal evolution}. The only simulations performed to study this mechanism to date indicate substantial cancellations \citep[][who also demonstrated the exact cancellation of this term in some simple stellar models]{BA2021}, such that this mechanism is unlikely to change the picture presented above. However, the effectiveness of this mechanism (and of other possible interactions between equilibrium tides and convection) is still debated and certainly warrants much further work, particularly by exploring more realistic models in -- albeit very challenging -- numerical simulations.

Equilibrium tide damping by turbulent convection is likely to be dominant in giant stars but is probably ineffective in most other tidal scenarios according to our current understanding. Unsolved questions include: how does realistic density variation affect the interaction with tidal flows and convection? Is Eq.~\ref{Deqm}, with $\nu_E(r)$, a suitable way to model the interaction between tides and convection in global models? Is $\mathcal{I}_2$ ever important and if so, when does it dominate over $\mathcal{I}_1$? What are the effects of magnetic fields, and can magnetic turbulent diffusion operate efficiently on equilibrium tides \citep[as hypothesised by][]{Wei2022}?

\subsubsection{Inertial waves}\label{IW}

\begin{figure}[b]
\vspace{-0.8cm}
\centering
\includegraphics[width=0.32\textwidth,trim=6cm 8.7cm 19cm 0cm,clip=true]{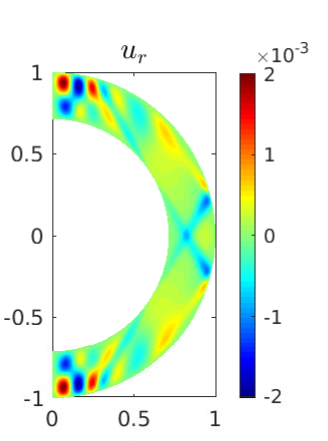}\qquad\qquad 
\includegraphics[width=0.4\textwidth,trim=6cm 0cm 8cm 0cm,clip=true]{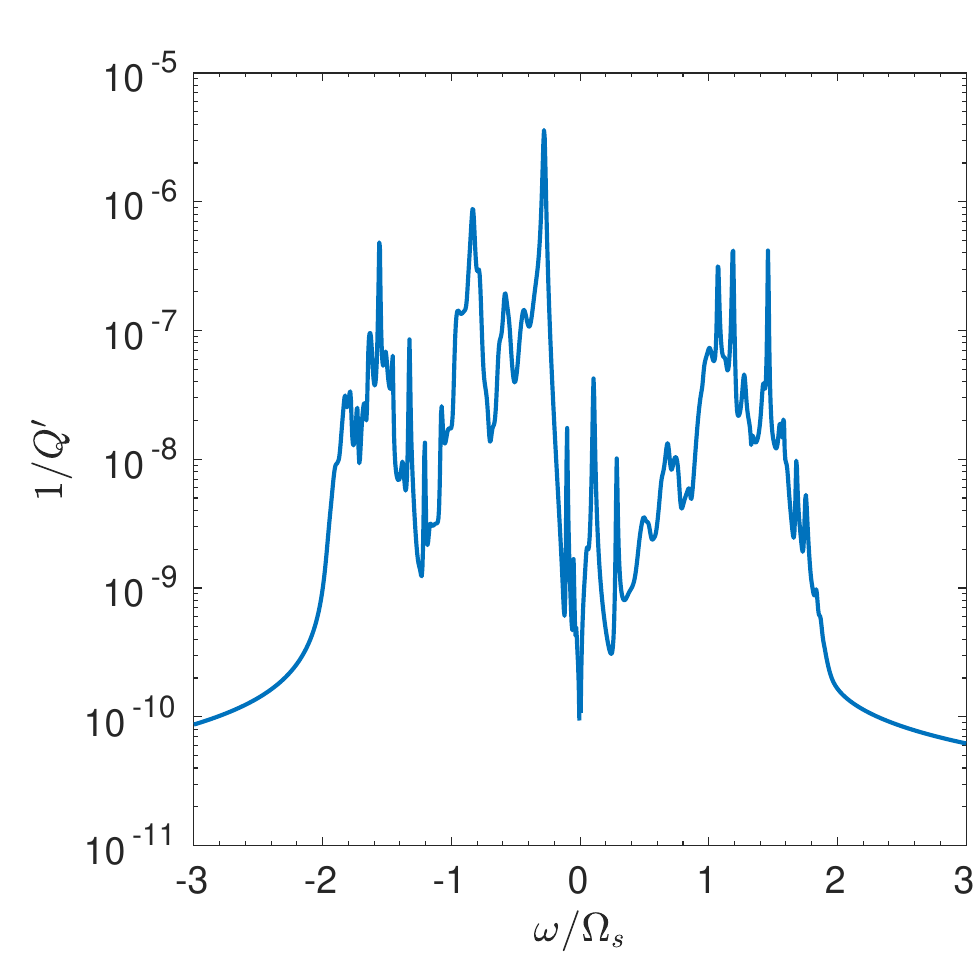}
\caption{Left: illustration of (linear) tidally excited inertial waves in a uniformly-rotating (anelastic) model of the solar convective envelope (with viscosity $\nu/(2\Omega_s R_1^2)=10^{-5}$), showing radial velocity (in units $\epsilon_T R_1\Omega_s$) in the meridional plane (the rotation axis lies along the $y$-axis) forced with $\omega=\Omega_s$. Right: inverse tidal quality factor ($1/Q'\propto D$) of inertial waves as a function of tidal frequency in the same model but with a rotation period of 10 days, showing the substantial enhancement within $|\omega|\leq 2|\Omega_s|$ when inertial waves are excited.}
\label{chap1:fig5}
\vspace{-1cm}
\end{figure}

Convection zones of rotating stars and planets are likely to be approximately neutrally stratified with $N^2\approx 0$ (to a first approximation) if convection is efficient in mixing entropy, so they support the existence of inertial waves restored by Coriolis forces. See the left panel of Fig.~\ref{chap1:fig5} for a visualisation of these waves in a solar model. These waves can be linearly\footnote{They can also be excited nonlinearly by sufficiently strong tides via fluid instabilities of the equilibrium tide in rotating bodies, namely the elliptical and precessional instabilities \citep[e.g.][]{LeBars2015,B2016,dVBH2023}.} excited by tidal forcing if $|\omega|\leq 2|\Omega_s|$. This, for example, is always relevant on a weakly eccentric ($e\ll 1$) orbit with a synchronised (and aligned) spin, since the only tidal frequencies are then $|\omega|=|\Omega_s|$. Hence, considering  inertial waves is relevant for orbital circularisation of fully convective stars or those with convective envelopes. It is also commonly satisfied in spin-orbit synchronisation and alignment of stellar binary stars, but it is often \textit{not} satisfied for observed hot Jupiter systems orbiting (aligned) slowly rotating stars, where these waves cannot be excited in the host stars.

Studies of the roles of inertial waves for tidal dissipation in planets and stars are relatively recent owing to the complexity of incorporating rotation when studying the tidal response \citep[early works studying them include:][]{SavPap1997,OL2004, Wu2005, PapIv2005,OL2007,GoodmanLackner2009,PapIv2010,RV2010}. In most cases numerical calculations  of the tidal response are required even in linear theory. Such calculations indicate these waves provide efficient dissipation for frequencies $|\omega|\leq 2|\Omega_s|$ (see Fig.~\ref{chap1:fig5}, right panel), substantially enhancing it over the case of a non-rotating fluid. The dissipation of inertial waves is predicted to scale as $D\propto Q'^{-1}\propto \Omega_s^2$ in linear theory, and hence, all other things being equal, rapidly rotating bodies tend to be more dissipative than slowly rotating ones when these waves are excited. A variety of models have been considered in prior work: polytropic models with or without a solid inner core \citep[e.g.][]{OL2004,PapIv2005}, incompressible envelopes \citep[e.g.][]{RV2010}, and some realistic stellar models \citep[e.g.][]{SavPap1997,OL2007,PapIv2010}. 

Calculations modelling convective envelopes indicate that these waves substantially enhance tidal dissipation in a complicated, strongly frequency-dependent manner, also depending on the fluid viscosity (more precisely, the Ekman number, the ratio of viscous to Coriolis forces). There is also a strong dependence on the geometry of the convective region, with full sphere geometry (fully convective low-mass stars) leading to excitation of global regular inertial modes, and spherical shells leading to excitation of inertial wave shear layers from the inner shell, with the latter being strongly enhanced for larger inner shells, particularly in incompressible models, where $D\propto Q'^{-1}\propto \alpha^{5}$ in some regimes ($\alpha$ is the ratio of the inner to outer radii of the convective envelope). However, the typical level of dissipation in any given configuration, using a frequency-integrated measure over the interval $-2|\Omega_s|\leq \omega \leq 2|\Omega_s|$ is much less sensitive to viscosity (i.e.~the taller peaks and deeper troughs, plus having more features for smaller viscosities, tend to cancel each other out) and may be largely independent of the damping mechanism.

For bound orbits, the periodic dissipative response is usually calculated by assuming a steady-state balance between viscous dissipation and tidal forcing under the assumption that convective turbulence damps inertial waves just like it can damp equilibrium tides (as in \S~\ref{EqmDamp}), by acting as a turbulent viscosity, much larger than -- but acting in a similar way to -- the molecular viscosity of the fluid. The validity of this approach has not yet been demonstrated though. It is also likely that magnetic fields modify the waves to become magneto-inertial, then magnetic diffusion can act upon these waves and more efficiently damp them \citep{LO2018,Astoul2019}. For tidal capture problems, i.e., parabolic encounters, only the excitation of waves upon periastron passage -- that are assumed to be subsequently fully dissipated by whatever mechanism -- is required, so there is less sensitivity to the nature of the damping mechanism.

Since the particular dissipative peaks and troughs in the tidal response are strongly sensitive to fluid (turbulent) viscosity and to the geometric structure of the convection zone, are also modified by magnetic fields \citep{LO2018}, nonlinear effects and differential rotation \citep{FBBO2014,AB2022,AB2023}, which each have significant uncertainties associated with them, there is a strong motivation to consider simplified approaches for astrophysical modelling. One approach is that of \citet{Ogilvie2013}, who presented a novel way -- using an impulsive forcing, which may be rigorously applied to tidal encounters, i.e., parabolic orbits -- to determine a certain (low-frequency) frequency-averaged measure of the dissipation of inertial waves (specifically $\int \mathrm{Im}\{k_l^m(\omega)\}\mathrm{d} \ln \omega$) that is independent of the specific damping mechanism, and captures the most important effects of stellar structure and the overall properties of inertial waves. This is equivalent to a constant (frequency-independent) $Q'$ representing the typical level of dissipation when inertial waves are excited, but with its value determined from the stellar structure and rotation. It was first evaluated in stellar models with a piece-wise homogeneous two-layer structure (i.e.~a homogeneous convective envelope on top of a homogeneous radiative core) by \citet{Mathis2015}, and this model has since been more widely adopted due to its simplicity. It has also been evaluated in realistic stellar models \citep{B20,B22}, which indicate that in low-mass and solar-type stars on the main-sequence, 
\begin{align}
Q' \approx 10^7 \left(\frac{P_{s}}{10 \mathrm{d}}\right)^2,
\end{align}
where $P_s$ is the stellar rotation period. PMS stars are more dissipative, so this stage may be crucial for solar-type binary circularisation.
This mechanism is likely to be the dominant one for explaining orbital circularisation and spin synchronisation of solar-type binaries, tidal circularisation and synchronisation of gaseous giant planets (hot and warm Jupiters) and for explaining the orbital migration of the moons of Jupiter \& Saturn in our solar system. It is therefore worthy of much further work.

Unsolved questions: how are these waves dissipated in stars? Are they dissipated through interaction with magnetic fields, turbulent convection, or nonlinearity? How do they interact with turbulent convection (is it reasonable to model this using a turbulent viscosity?)?  How are they excited and how do they propagate and dissipate in differentially rotating and magnetised bodies? Can resonant locking occur for these waves \citep[e.g.,][]{Wu2024}? How does non-sphericity of stars due to rotational and (large amplitude) tidal deformations  affect the tidal response \citep[e.g.~][]{B2016,Dewberry2023}?

\subsection{Radiative zones}

Radiative zones are regions where heat is transported by radiative processes rather than by fluid motions, and they have $N^2>0$, such that buoyancy forces are stabilising and gravity waves are supported. Since these regions are convectively stable, it is often thought that they are quiescent regions, though this is an oversimplification. Without tides, radiative zones host a spectrum of convectively-excited gravity waves, and fluid motions such as meridional circulations and those driven by fluid instabilities (e.g.~due to differential rotation, or double diffusion if there are competing compositional and thermal gradients). However, the fluid motions in radiative layers are thought to be much weaker than those operating in convective layers, so we do not often consider tidal flows to be damped by a turbulent viscosity (or diffusivity) like in \S~\ref{EqmDamp}. The dominant linear hydrodynamic damping mechanism acting on tidal flows is radiative diffusion of their associated temperature perturbations, which is generally unimportant for equilibrium tides \citep[e.g.][]{Zahn1977}, hence the focus is usually on dynamical tide dissipation mechanisms in radiative zones\footnote{Equilibrium tidal flows in radiative zones can also be unstable to parametric instabilities exciting pairs of daughter gravity (or gravito-inertial) waves much like the elliptical instability for large tidal amplitudes \citep[e.g.][]{Weinberg2012}.}.

\subsubsection{Internal gravity waves (or g-modes)}\label{IGW}

\begin{figure}[t]
\quad 
\includegraphics[width=0.47\textwidth,trim=0cm 0.93cm 0cm 0cm,clip=true]{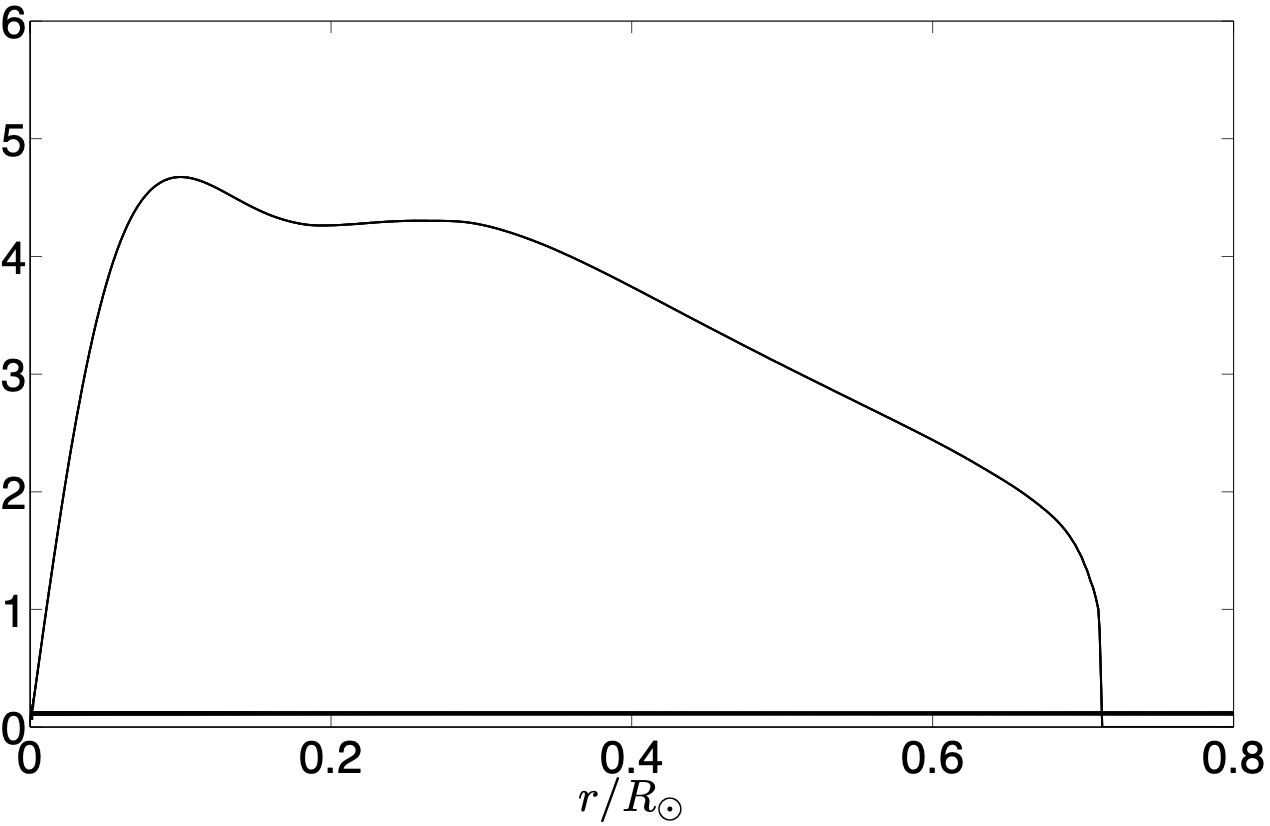}\qquad
\includegraphics[width=0.47\textwidth,trim=3.5cm 6.6cm 3cm 0.45cm,clip=true]{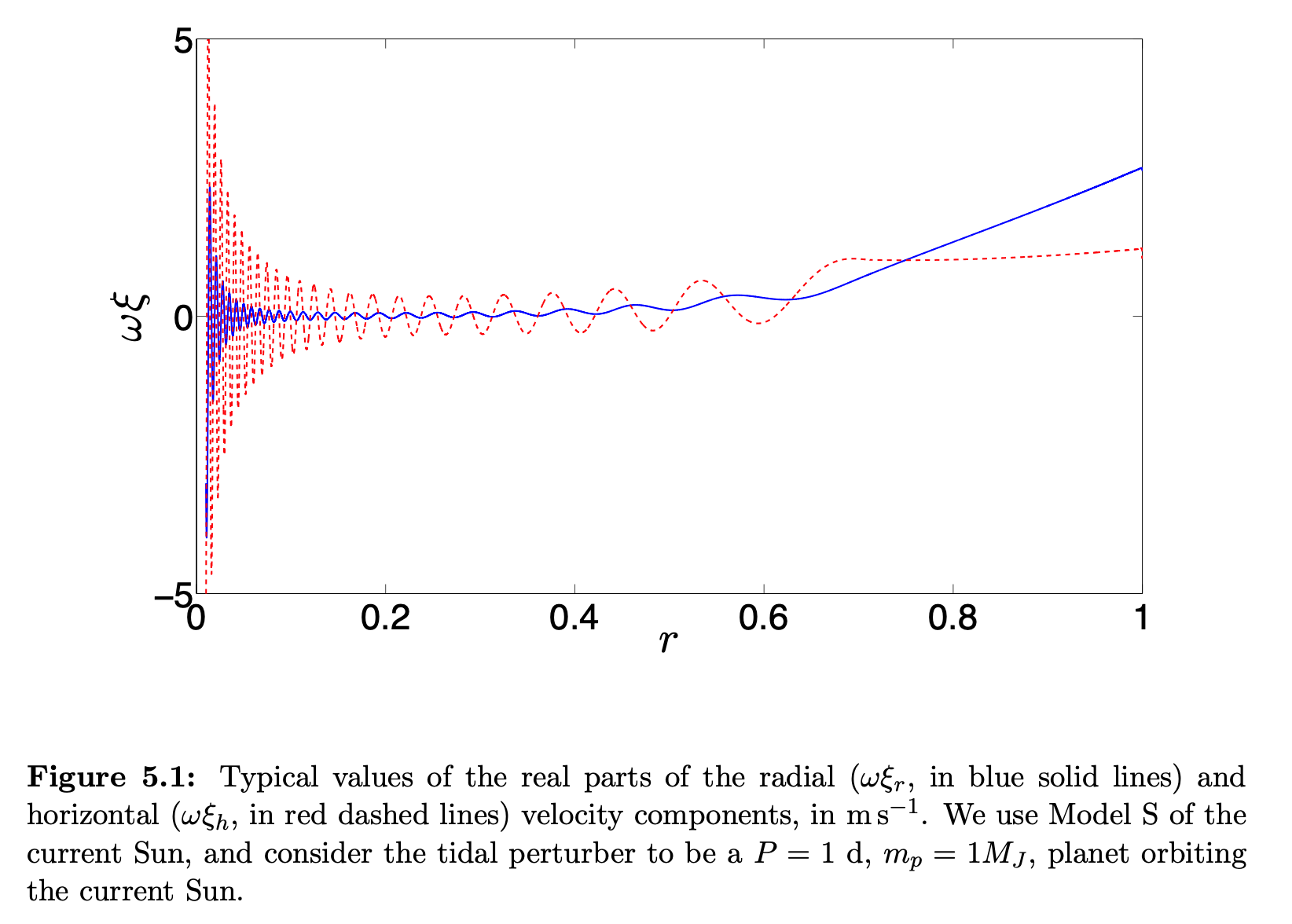}
  \begin{picture}(0,0)
        \put(-340,-2){$r/R_\odot$}
        \put(-120,-2){$r/R_\odot$}
        \put(-225,138){$u_r, u_h \, (m/s)$}
        \put(-485,80){$N/\omega_d$}
    \end{picture}\\
\caption{Left: plot of $N/\omega_d$ vs radius in a model of the Sun \citep[model S from][]{JCD1996}, with the tidal frequency for a 1 day orbit indicated by the horizontal line \citep[modified from Fig.~1 of][]{BO2010}. Right: tidal response of a slowly rotating Sun to an orbiting companion on a 1 day orbit (0.5 day tidal period), plotting radial velocity (blue solid) and horizontal velocity (red dashed) amplitudes in $m/s$ as a function of fractional radius  \citep[modified from Fig.~7 in][]{B2011}. This shows the smoothly-varying equilibrium tide in the convective envelope ($r\gtrsim 0.71R_\odot$), below which the response is primarily oscillatory ($r\lesssim 0.71R_\odot)$ due to launching of gravity waves inwards from the radiative/convective interface.}
\label{chap1:fig6}
\end{figure}

Radiation zones are stably stratified and thus support gravity waves -- or standing g-mode oscillations -- which can be resonantly excited by tidal forcing if tidal frequencies satisfy $0\leq |\omega|\leq N$ (for slow stellar rotation, otherwise both limits are modified and the waves are gravito-inertial), which is very often satisfied in applications (e.g.~left panel of Fig.~\ref{chap1:fig6}). These waves are primarily excited at the radiative/convective interface and propagate from their launching sites through the radiation zone until they reflect from boundaries and set up standing mode oscillations or they are dissipated. They were first studied in simple polytropic stellar models (in which $p\propto \rho^{1+1/n}$ for a given index $n$) by \citet{Cowling1941} and then for their tidal response in massive (``early-type") stars with radiative envelopes starting with \citet{Zahn1970,Zahn1975,Zahn1977}, then in solar-type stars with radiative cores starting with \citet{GD1998,T1998}. 

Radiative diffusion is a particularly efficient damping mechanism for short-wavelength gravity waves near the surfaces of massive stars  \citep[e.g.~][]{Zahn1975,GN1989}, but tends to be weaker in radiative cores. If damping mechanisms are efficient, tidal dissipation can be computed assuming travelling waves launched from the radiative/convective interface are fully damped before they reflect to form standing modes. The resulting dissipation is a smoothly-varying function of tidal frequency that can be simply computed in stellar models. On the other hand, if damping mechanisms are weak, tidal responses are dominated by tall, narrow (in frequency) resonant peaks, where tidal forcing is resonant with a free gravity mode oscillation. Naively, efficient dissipation would only then be expected when the system is lucky enough to be in resonance. Since there are many resonances though, it is certain that many of these will be passed through in the lifetime of a system. ``Resonance locking" has also been proposed, where the joint evolution of the tidal frequency and the g-modes is such that resonance is maintained, resulting in sustained enhanced dissipation \citep[e.g.][]{WS1999,MaFuller2021}.

Since gravity waves typically have much shorter wavelengths than the stellar radius, nonlinear effects acting on these waves can be more important than for equilibrium tidal flows. Fig.~\ref{chap1:fig6} shows an example calculation of the linear tidal response (with $l=m=2$) in a slowly-rotating solar-type star hosting a hot Jupiter planet on a 1 day circular orbit. We observe a non-oscillatory equilibrium tide response in the convection zone and an oscillatory wave-like response in the radiative zone, corresponding with propagating gravity waves (due to the absorbing inner boundary adopted). The wavelengths of these waves decrease as they approach the centre partly because $N$ increases inwards throughout most of the radiation zone. The wave amplitudes can become quite large as they approach the centre of the star due to geometrical focussing (i.e.~wave energy in each wavelength becomes concentrated into an increasingly smaller volume as $r\to 0$, hence the wave amplitude must go up). In fact, the amplitude can become sufficiently large that these waves are strongly unstable, undergoing wave breaking analogous to surface waves on the ocean breaking as they approach the shore, but in this case the surface that overturns is an internal stratification surface (of constant entropy), and the instability is primarily convective in nature. This happens when the crest of the wave overtakes the trough, and is therefore strongly nonlinear, which can be estimated using \citep[e.g.][]{GD1998}
\begin{align}
\xi_r k_r \gtrsim 1,
\end{align}
where $\xi_r$ is the radial displacement (of the stratification surfaces) in the wave and $k_r$ is the radial wavenumber. When wave breaking occurs, waves deposit angular momentum in the fluid locally, which can spin up or down portions of the star to match the orbital frequency \citep{GN1989,BO2010,B2011}. For weaker wave amplitudes, gravity waves are known to be unstable to parametric instabilities where the wave interacts with pairs (or more) of daughter waves (the simple case of a plane propagating ideal gravity wave is unstable in this way for any non-zero wave amplitude), draining energy from the primary wave and leading to enhanced tidal dissipation \citep[e.g.][]{BO2011,Weinberg2012,Weinberg2024}. Waves also deposit momentum and modify the rotation profile of the star as they are weakly dissipated by radiative diffusion \citep[e.g.][]{Guo2023}, and this differential rotation can back-react on the waves and modify their properties. Nonlinear effects may impede resonance locking from operating in solar-type stars \citep[e.g.][]{Guo2023}, though the possibility of resonance locking in other stars should be explored further \citep[e.g.][]{MaFuller2021}.

When gravity waves are launched from radiative/convective interfaces and are in the ``travelling wave" regime, being fully damped by whatever process (radiative damping, wave breaking, other nonlinear wave-wave interactions, conversion to magnetic waves etc.), the resulting tidal dissipation can be computed simply in the low-frequency regime using stellar models. For solar-type stars, we typically find \citep{GD1998,OL2007,B2011,IvPapCh2013,B20,Ahuir2021}
\begin{align}
Q'\approx 10^5 \left(\frac{P_{\mathrm{tide}}}{0.5 \; \mathrm{d}}\right)^{\frac{8}{3}},
\end{align}
where $P_\mathrm{tide}=2\pi/\omega$ is the tidal period \citep[see Fig.~8 of][for an exploration of different stellar masses and ages]{B20}. Note that this regime is independent of the details of the damping mechanism as long as the waves are fully damped inside the star. This regime can also be calculated in massive (early-type) stars \citep[e.g.][]{Zahn1975}. Predictions are modified for larger tidal frequencies \citep[e.g.][]{B2011,IvPapCh2013}, where standing modes may be more important \citep{MaFuller2023}, and by sufficiently rapid rotation.

The effects of magnetic fields on gravity waves have been poorly explored to date and are a promising avenue of further research. One possibility for stars with convective cores on the main sequence -- for which wave breaking is less likely as the waves cannot approach the centre of the star -- is for tidal waves to be converted into magnetic waves when they propagate into a region with a sufficiently strong magnetic field near the stellar core \citep{DdVLB2024}. This idea follows earlier suggestions that this process could be important for oscillations of red giants \citep[e.g.][]{Fuller2015}.

Unsolved questions include: how do magnetic fields affect the excitation, stability and evolution of these waves? How do these waves evolve nonlinearly and how do they interact with differential rotation (these issues are only partly understood)? How do the separate compositional and thermal stratifications (and diffusivities) affect the evolution of these waves? Is resonance locking of these waves a viable tidal dissipation scenario? The latter is an interesting idea with an attractive simplicity worth exploring further, but whether or not it can work in practice is unclear at present.

\section{How do we know tides are important? Some observational evidence}
\label{Obs}
\begin{figure}[b]
\includegraphics[width=0.3\textwidth,trim=0.1cm 0cm 0cm 0cm,clip=true]{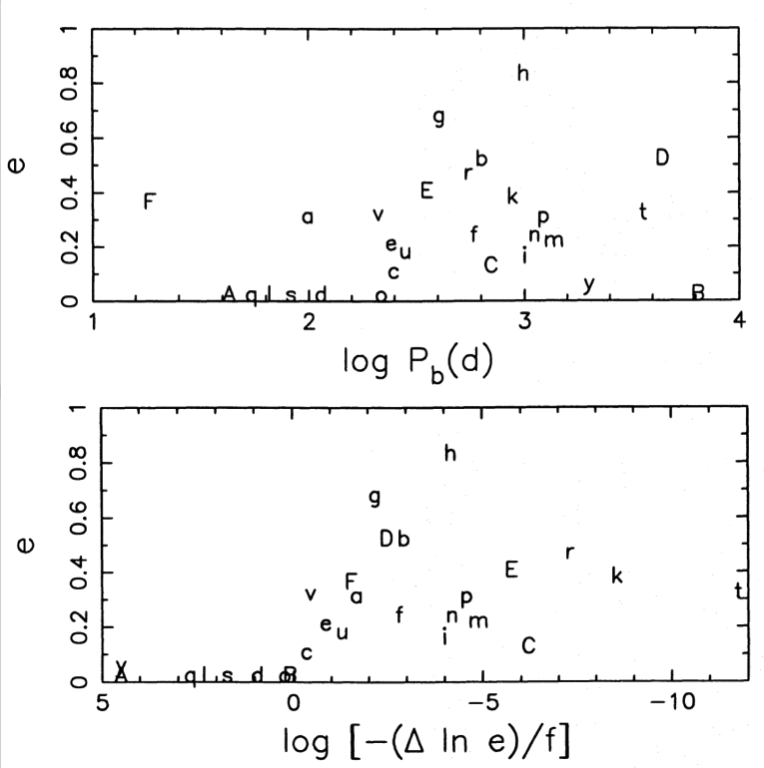}\quad
\includegraphics[width=0.3\textwidth,trim=0cm 0cm 0cm 0cm,clip=true]{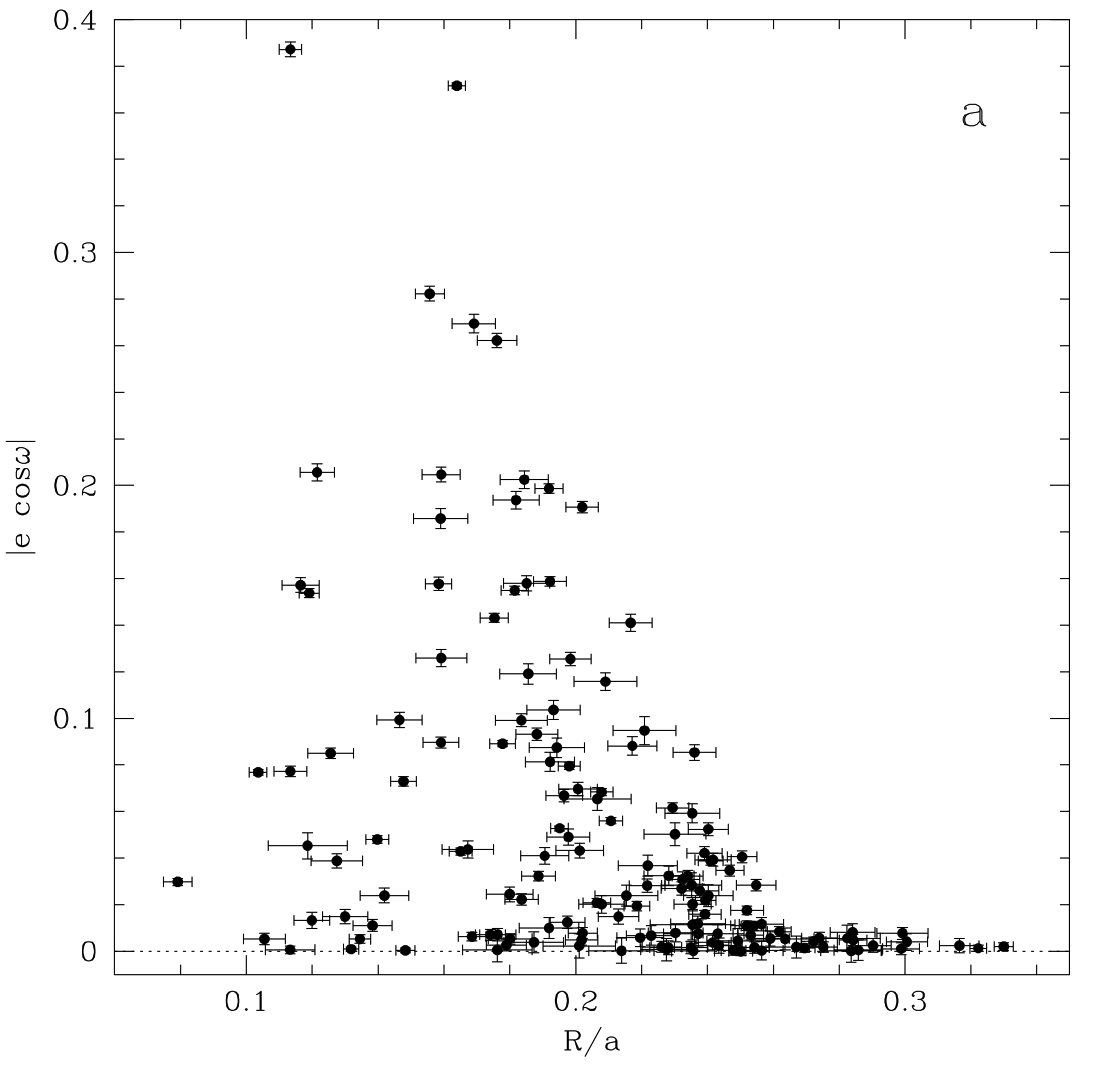}\quad
\includegraphics[width=0.35\textwidth,trim=0cm 1.3cm 0cm 0cm,clip=true]{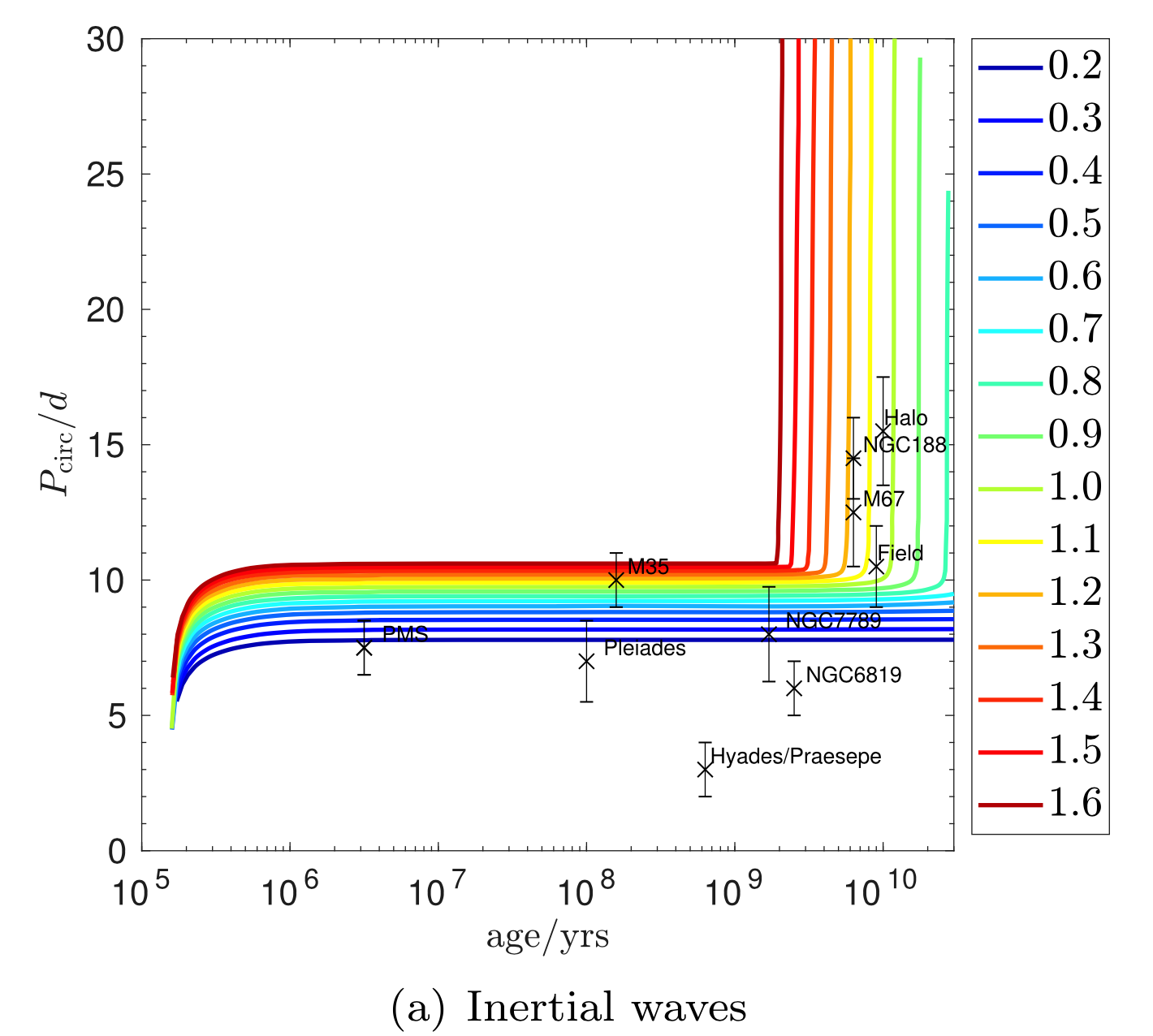}
\caption{Left: eccentricities ($e$) of spectroscopic binaries containing a giant star as a function of the base 10 logarithm of orbital period ($P$) in days in the top panel and the predicted change in eccentricity due to dissipation of equilibrium tides in the bottom panel, from Fig.~4 in \citet{VP1995}, indicating predicted complete circularisation when the $x$-axis value is 0. [Credit: Verbunt \& Phinney, A\&A, 296, 709, 1995, reproduced with permission \copyright ESO]. Middle: eccentricities of early-type (massive) eclipsing star binaries with radiative envelopes in the Small Magellanic Cloud from Fig.~1 in \citet{NorthZahn2003}, noting that theory of gravity wave dissipation predicts circularisation for $R/a\gtrsim 0.25$. [Credit: North \& Zahn, A\&A, 405, 677, 2003, reproduced with permission \copyright ESO] Observations are in very good agreement with theory in these two examples. Right: circularisation periods ($P_\mathrm{circ}$) of solar-type binary stars as a function of stellar age, comparing observations from various stellar clusters in \citet{Nine2020} with theory of inertial wave dissipation (for various stellar masses, indicated by colours) in the PMS and later on the MS \citep[][the latter MS evolution is not overly sensitive to uncertain initial conditions when tides begin operating, but PMS evolution is, and predictions can be shifted to match observations for each mass on the PMS if desired]{B22}.}
\label{chap1:fig7}
\end{figure}

There is substantial observational evidence for (dissipative) tidal evolutionary processes sculpting the properties of stellar and planetary systems. Some (but by no means all!) of this wealth of evidence is briefly summarised below -- with pointers to the wider literature -- including the extent to which current tidal theories work to explain each aspect. Much evidence relates to eccentricity distributions of binary stars of various spectral types. These generally show that the closest binaries exhibiting the strongest tidal interactions tend to be approximately circular, likely caused by tidal dissipation inside these stars, whereas wider binaries experiencing weaker tides tend to be eccentric (presumably as a result of their formation). Some selected examples are plotted in Fig.~\ref{chap1:fig7}. There is also evidence of tidal evolution of stellar spins (rotations) towards synchronism with the orbit, and for orbital migration of hot Jupiters.

\begin{itemize}
\item \textbf{Early-type stars with radiative envelopes:} Circularisation of the orbits, and synchronisation of the spins, of early-type massive main sequence stars by gravity waves (assuming the fully damped regime discussed in \S~\ref{IGW}) in radiative envelopes appears to match observations well \citep[see the middle panel of Fig.~\ref{chap1:fig7} here and][]{NorthZahn2003,KK2010}, therefore largely validating the theory of \citet{Zahn1975} in these stars. However, there are significant outliers motivating further theoretical work \citep[e.g.][]{Albrecht2021}. The evolution of stellar spin for stars in eccentric orbits is not well explained or understood \citep[e.g.][]{Zimmerman2017}. 
\item \textbf{Red giant stars:} \citet{VP1995} \citep[and more recently][]{PW2018,Beck2018} applied equilibrium tide damping (along with stellar evolution) using the simplest turbulent viscosity from mixing-length theory \citep[with $\nu_E=u_cl_c/3$;][]{Zahn1989} outlined in \S~\ref{EqmDamp} to explain the circularisation of spectroscopic binaries containing a giant star. These stars are not in the fast tides regime, so the frequency-reduction of the turbulent viscosity for fast tides can largely be ignored, and these results (shown in the left panel of Fig.~\ref{chap1:fig7}) appear to largely validate the theory of equilibrium tide damping in giant stars (in the slow tides regime).
\item \textbf{Solar-type and low-mass stars with convective envelopes:} Circularisation periods of solar-type binaries in clusters with different ages \citep[e.g.][]{Meibom2005,Nine2020} are observed to depend on age, and indicate both efficient PMS dissipation and processes operating later on the MS. Tidal excitation and dissipation of inertial waves (using the frequency-averaged measure described in \S~\ref{IW}) in the PMS phase, and later on the MS for nearly spin-synchronised binaries, can explain circularisation periods 
of solar-type binaries during the PMS \citep[as required by][]{Bashi2023} and also later on the MS \citep[see right panel of Fig.~\ref{chap1:fig7} here and][]{B22}. Earlier pioneering work by \citet{ZahnBouchet1989} identified the importance of the PMS phase for tidal evolution in their models assuming equilibrium tide damping, but their calculations ignored its inhibition in the fast tides regime (see \S~\ref{EqmDamp}). Eccentricity distributions of larger populations of low-mass and solar-type stars have also been studied and compared with aspects of tidal theory \citep[e.g.][]{VanEylen2016,Triaud2017,Albrecht2021,Zanazzi2022,Penev2022,Bashi2023}, particularly using photometric binaries discovered with space missions such as Kepler and TESS. These provide important information on both tidal evolution and the formation scenarios of close binaries, with many aspects that have yet to be fully compared with (or explained by) tidal theory. Evidence for tidal spin synchronisation of approximately solar-type binaries \citep{Meibom2006,Lurie2017,Patel2022} have also not been fully explained by tidal theory to date -- and are likely complicated by (latitudinal) differential rotation in stars with convective envelopes, preventing perfect synchronism from being achieved.
\item \textbf{Tidal spin-up of exoplanet host stars and orbital migration/destruction of hot Jupiters:} Most hot Jupiter host stars rotate much slower than their planets orbit them, so stellar tidal dissipation drives these planets to spiral inwards and, perhaps, to ultimately be destroyed, while the stars are spun up in the process \citep[evidence for the latter includes, e.g.,][]{Maxted2015,Penev2018,Ilic2024}. The statistical analysis by \citet{CC2018} finds stars in which inertial waves can be excited to have $Q'$ consistent with the theory outlined in \S~\ref{IW} \citep{B20}. On the other hand, the results of \citet{Penev2018} indicating a frequency-dependent $Q'$ that sharply increases at short tidal periods has not been explained. There is also evidence from statistical analyses of the population of hot Jupiters that some of these planets are destroyed as their stars evolve on the main sequence \citep[e.g.][]{HS2019}. We are now in an exciting age where slight deviations in the orbital period of a planet (as small as tens of ms$/$year) can be inferred based on comparing predicted and observed times of transit (when the planet passes in front of the star and causes a dimming in starlight) over many years, or a decade or more \citep[particularly WASP-12 b, e.g.,][]{Maciejewski2016,Patra2020,Turner2021}. This is potentially caused by tidally-driven orbital decay, and so these observations can directly test tidal theories (those of gravity wave dissipation described in \S~\ref{IGW} are likely to be particularly relevant here). 
\item \textbf{Circularisation of hot Jupiters:} Hot Jupiters have an eccentricity distribution indicating those with orbital periods longer than about 10 days are primarily eccentric, while those orbiting more closely are primarily circular or tend to have lower eccentricities \citep[e.g.][]{Jackson2008,Hansen2010,Hansen2012,Mahmud2023}. Tidal dissipation of inertial waves may be important in explaining observed trends \citep{OL2004,LBdVA2024}, but more work is required to explore this further.
\item \textbf{Spin-orbit alignment of binary stars and hot Jupiters:} Observations combining transits and radial velocity data can infer the sky-projected spin-orbit angle of some hot and warm Jupiters, and stellar binaries \citep[e.g.~for hot Jupiters,][]{Albrecht2012}. These are likely to be affected by tides, but it is unclear currently how much of the distribution is primordial, due to formation, and how much is due to later tidal evolution. It has been proposed that inertial waves could be excited in convective envelopes on misaligned orbits even if the star spins slowly \citep{Lai2012,LO2017}, which may contribute to observed trends, but more theoretical work is required on this problem \citep[gravity modes have also been proposed, e.g.,][]{Zanazzi2024}. 
\item \textbf{Evolution of Jupiter's and Saturn's moons:} Astrometric observations over more than a century of the positions of the moons of Jupiter and Saturn indicate efficient tidal dissipation in these planets \citep[e.g.][]{Lainey2009,Lainey2012}. Similar to the Earth-Moon system, these planets rotate faster than their moons orbit them, so tides drive the moons to generally migrate outwards. This can be explained by dissipation of inertial waves in convective regions of these planets, with contributions from gravito-inertial waves in the deeper stably-stratified fluid layers that have recently been inferred for these planets \citep{Lin2023,Dewberry2023,Pontin2023,Dhouib2024}. Resonance locking \citep{Fuller2016} and efficient equilibrium tide dissipation have also been proposed \citep{T2021}.
\end{itemize}

There are various challenges in applying tidal theory to explain observations. Firstly, the initial conditions before tides are applied must be known, since tidal evolutionary timescales are strongly dependent on orbital period (for example), which typically requires a good understanding of the formation scenarios of binaries and planetary systems -- which may itself involve tidal dissipation (e.g.~high-eccentricity migration of hot Jupiters and close binaries). Secondly, tidal dissipation efficiencies (i.e. $Q'$) are known to vary substantially with tidal frequency (period), stellar mass and age, and with tidal amplitude, so a study must not assume all stars (and planets) to have the same $Q'$ to obtain meaningful results. Thirdly, in some cases coupled evolution of the stellar structure and tidal evolution are required (e.g.~in PMS stars or red giants) and using tidal timescales is not always sufficient. Fourthly, there are still remaining uncertainties in many aspects of tidal theory, but substantial progress is being made. Future work should incorporate sophisticated tidal models in dynamical models of multiple star systems and planet formation, and in stellar and planetary population synthesis codes.

\section{Conclusion and Future Outlook}
This article is intended to provide an introduction to tidal flows and dissipation in stars and planets, and the roles they play in driving spin and orbital evolution in stellar multiples and planetary systems. We began by introducing the tidal potential and its properties, then provided two simple illustrations of how dissipation of tidal flows drives spin and orbital evolution, before introducing tidal responses in stars and describing their decomposition into equilibrium/non-wavelike and dynamical/wavelike tides. A brief summary of current theoretical understanding of the mechanisms by which tidal flows are dissipated was presented, before reviewing some observational evidence indicating the important role of tidal evolution in stellar and planetary systems. This is an exciting field, expected to be revolutionised by both theoretical advances and new observational constraints, including those expected with space missions such as PLATO, over the next few years.

\section{Further information}
Some excellent and helpful review articles in recent years on tides in stars and planets, containing much further information, include:
\begin{itemize}
\item \citet{Ogilvie2014}
\item \citet{Mathis2019}
\item \citet{Fulleretal2024}
\item ``Main-sequence exoplanet systems: tidal evolution" by K.~Penev in `Encyclopedia of Astrophysics' (Editor-in-Chief: Ilya Mandel, Section Editor: Dimitri Veras)
\item ``Giant branch planetary systems: Dynamical and radiative evolution" by A.~Mustill in `Encyclopedia of Astrophysics' (Editor-in-Chief: Ilya Mandel, Section Editor: Dimitri Veras)

\end{itemize}

\begin{ack}[Acknowledgments]
\noindent
AJB was supported by STFC grant ST/W000873/1. I would like to thank the section editor, Fabian Schneider, for very helpful suggestions that have improved this article.
\end{ack}

\bibliographystyle{Harvard}
\bibliography{reference}

\end{document}